\documentclass[a4paper]{article}

\usepackage[english]{babel}
\usepackage[utf8x]{inputenc}
\usepackage[T1]{fontenc}

\usepackage[a4paper,top=3cm,bottom=2cm,left=3cm,right=3cm,marginparwidth=1.75cm]{geometry}

\usepackage{amsmath}
\usepackage{amssymb}
\usepackage{graphicx}
\usepackage[colorinlistoftodos]{todonotes}
\usepackage[colorlinks=true, allcolors=blue]{hyperref}

\title{Edge-based compartmental modelling of an SIR epidemic on a dual-layer static-dynamic multiplex network with tunable clustering}
\author{Rosanna C Barnard\textsuperscript{1,2*},
Istvan Z Kiss\textsuperscript{1},
Luc Berthouze\textsuperscript{2},
Joel C Miller\textsuperscript{3}}

\begin{document}
\maketitle

\begin{flushleft}
\textbf{1} Department of Mathematics, University of Sussex, Falmer, Brighton, United Kingdom
\\
\textbf{2} Centre for Computational Neuroscience and Robotics, University of Sussex, Falmer, Brighton, United Kingdom
\\
\textbf{3} Institute for Disease Modeling, Bellevue, WA, United States of America \\ \bigskip * rosannabarnardresearch@gmail.com \end{flushleft}

\begin{abstract}
The duration, type and structure of connections between individuals in real-world populations play a crucial role in how diseases invade and spread. Here, we incorporate the aforementioned heterogeneities into a model by considering a dual-layer static-dynamic multiplex network. The static network layer affords tunable clustering and describes an individual's permanent community structure. The dynamic network layer describes the transient connections an individual makes with members of the wider population by imposing constant edge rewiring. We follow the edge-based compartmental modelling approach to derive equations describing the evolution of a susceptible - infected - recovered (SIR) epidemic spreading through this multiplex network of individuals. We derive the basic reproduction number, measuring the expected number of new infectious cases caused by a single infectious individual in an otherwise susceptible population. We validate model equations by showing convergence to pre-existing edge-based compartmental model equations in limiting cases and by comparison with stochastically simulated epidemics. We explore the effects of altering model parameters and multiplex network attributes on resultant epidemic dynamics. We validate the basic reproduction number by plotting its value against associated final epidemic sizes measured from simulation and predicted by model equations for a number of setups. Further, we explore the effect of varying individual model parameters on the basic reproduction number. We conclude with a discussion of the significance and interpretation of the model and its relation to existing research literature. We highlight intrinsic limitations and potential extensions of the present model and outline future research considerations, both experimental and theoretical.
\end{abstract}

\section*{Introduction}
The continual design and development of mathematical models describing epidemic processes on large, complex populations improves our understanding of how diseases and individuals behave during an epidemic, and how preventative measures can be implemented for the greater good. With ever-increasing computational power, models can incorporate increasingly complex features, and model predictions may become more valuable. Nonetheless, any model must tread a careful balance between capturing observed real-world complexity and enabling calculations and conclusions to be drawn with ease. The ultimate epidemiological model must therefore incorporate the behavioural and structural features which significantly influence disease dynamics, whilst being analytically tractable.

Social heterogeneity describes the propensity for a social group to be diverse in character or content, and is an important determinant when studying the dynamics and control of infectious diseases \cite{arthur2017contact}. In a social group, heterogeneity encompasses many descriptive elements, such as variations in individuals' behaviour or in susceptibility across group members. In network theory, social heterogeneity can also describe variations in the \emph{types} of connections an individual makes. For example, an individual can be connected to other individuals in distinct groups, such as workplace or community groups. 

Structured populations with multiple connection types are well described by multiplex networks, where a population of individuals partakes in multiple network \emph{layers}. Each network layer describes a specific type of interaction between members of the population, and network structure in one layer is allowed overlap with network structure in another layer. A pair of individuals in a multiplex network can share more than one connection. In a multiplex network, an individual is present in every network layer, but may or may not partake in connections in individual network layers.

Existing multiplex modelling studies have shown that single-layer approximations or aggregations of multiplex networks are not accurate enough to describe the epidemic process \cite{diakonova2016irreducibility,zhuang2016clustering,gomez2013diffusion,cozzo2013contact}, and further that an epidemic can spread on a multiplex network even if the individual layers are well below their respective epidemic thresholds \cite{zhao2014multiple}. A global cascades model generalised for multiplex networks was used to show that multiplexes are more vulnerable to global cascades than single layer networks \cite{brummitt2012multiplexity}. These studies highlight the importance of accounting for heterogeneity in connection type by considering multiplex network models. 

Another determinant of infectious disease dynamics is heterogeneity in the structural connections between individuals, within a single type of connection. Real-world networks often exhibit community structure, with a high density of connections within communities and a low density of connections between communities. They are also considered to exhibit other structural characteristics such as network transitivity or clustering, described in social network theory as the propensity for an individual to be connected to a friend of a friend \cite{newman2003structure}. 

Community structure has been shown to affect disease dynamics on single-layered (uniplex) networks, where on average, epidemics occurring on networks with community structure exhibit greater variance in final epidemic size, a greater number of small, local outbreaks that do not develop into epidemics, and higher variance in the duration of the epidemic \cite{salathe2010dynamics}. Network quality functions able to detect community structure in multiplex networks have been developed \cite{mucha2010community}. Further, results such as the large graph limit of an SIR epidemic process on a dynamic multilayer network, where one network layer represents community links and another represents connections in healthcare settings, have been derived \cite{jacobsen2016large}.

In network models, increased clustering is generally considered to slow an epidemic by increasing the epidemic threshold \cite{miller2009percolation}. However, this relationship is not always monotonic. Higher clustering in a multiplex study of information propagation led to an increase in the epidemic threshold and a decrease in final epidemic size \cite{zhuang2016information}. Increased clustering in a study of Watt's threshold model generalised for a multiplex network comprised of clustered network layers led to a decrease in the probability of a global cascade and its size \cite{zhuang2016clustering}. However, the authors also discovered a critical threshold for the average degree, above which clustering was shown to \emph{facilitate} global cascades \cite{zhuang2016clustering}. A uniplex network study found that simultaneously increasing clustering and the variance of the degree distribution led to an increase in final epidemic size \cite{volz2011effects}. Moreover, clustering can lead to correlations where high-degree individuals are more likely to connect with other high-degree individuals. It is clear that the effect of clustering is complex and should be considered in the design of network models.

In epidemiology it is also important to consider heterogeneity across contact \emph{duration}. In human populations, links between individuals may be long-lasting (persistent), e.g. between an infant child and their caregiver; temporary (transient), e.g. between workplace colleagues; or more short-lived (fleeting), e.g. between strangers coming into close proximity on public transport. In a study using a year's mobile phone data as a proxy for the structure and dynamics of a large social network, researchers found that persistent links tend to be reciprocal and are more common for individuals with low degree and high clustering \cite{hidalgo2008dynamics}. Many network-based studies in the past have considered fully static network structures, and hence solely investigate the effects of persistent connections between individuals, see \cite{keeling2005networks} for a review of differing approaches. 

Later studies of epidemic processes on networks have incorporated persistent \emph{and} transient connections into their models by imposing rewiring rules on static networks. Rewiring rules considered include spatially-constrained rewiring \cite{rattana2014impact}, random link activation and deletion \cite{taylor2012epidemic,selley2015dynamic,kiss2012modelling}, and temporary link deactivation \cite{tunc2013epidemics,shkarayev2014epidemics}. On the other hand, epidemic processes with fleeting contact duration can be well-described via the mass action model, which assumes all pairs of individuals contact one another at the same rate, the mean-field social heterogeneity model (also known as the degree-based mean-field model), which generalises the mass action model by allowing for variations in contact rate across the population, and the dynamic fixed- and dynamic variable-degree models, where edges are swapped at a given rate, or edges are broken and created at given rates, respectively \cite{miller2013model,miller2012edge}.

Here, we suppose that static and dynamic connections \emph{coexist} in any complex population. We aim to derive a network model describing an SIR epidemic process spreading through a population where each individual has two types of connections: persistent links to individuals in their household, constituting a static network layer with community structure, and transient connections to strangers in the wider population, where all such edges rewire at a constant rate, constituting a dynamic network layer with conserved degrees. 

In what follows, we utilise the edge-based compartmental modelling (EBCM) approach \cite{volz2008sir,miller2011note,miller2014epidemics,miller2012edge}, deriving equations which describe the time evolution of classical quantities of interest, where the underlying dual-layered static-dynamic network has heterogeneity in contact-type, contact-duration, and contact-structure. We derive the associated basic reproduction number $R_0$, following the next generation matrix approach \cite{diekmann2009construction}. We describe the implementation of the EBCM model and of statistically-correct Gillespie simulations of the epidemic process \cite{gillespie1976general}. The new model is validated, firstly by showing that collapsing either the static or dynamic network layers leads model equations to converge to existing equivalent model equations, and secondly by comparing the dynamics predicted by model equations to those from exact simulations. We explore how various combinations of model parameters and network layers influence global dynamics, uncover behavioural regimes that the model can achieve for specific combinations of infection and rewiring rates, and show that our derived $R_0$ behaves as expected. The paper concludes with a discussion of potential implications of the work as well as possible extensions.

\section*{Methods}

Our solutions are based on the class of undirected random graphs (networks). Each node is a member of a random number of static lines (2-vertex cliques), static triangles (3-vertex cliques) and dynamic lines (2-vertex cliques). The probability that a node has $s$ static line stubs, $t$ static triangle \emph{corners} and $d$ dynamic line stubs is described by the probability mass function $p_{s,t,d}$. The model captures network structure using the probability generating function (PGF) \begin{equation} g(x,y,z)=\sum_{s,t,d}p_{s,t,d}x^{s}y^{t}z^{d}. \label{eq:pgf_g} \end{equation} When differentiating the PGF \eqref{eq:pgf_g}, we use superscripts such that $g^{(x)}$ denotes the first (partial) derivative of $g$ with respect to $x$ and $g^{(y,y)}$ denotes the second (partial) derivative of $g$ with respect to $y$. Equation \eqref{eq:pgf_g} can be used to calculate useful properties of the multiplex network. For example, $M$, the expected number of static line stubs that belong to a randomly selected individual, $\hat{M}$, the expected number of static triangle corners that belong to a randomly selected individual, and $\tilde{M}$, the expected number of dynamic line stubs that belong to a randomly selected individual, are calculated as follows: \begin{align*} M &= \sum_{s,t,d}s p_{s,t,d}=g^{(x)}(1,1,1), \\ \hat{M} &= \sum_{s,t,d}t p_{s,t,d} = g^{(y)}(1,1,1), \\ \tilde{M} &= \sum_{s,t,d} d p_{s,t,d} = g^{(z)}(1,1,1). \end{align*} 

We consider a basic SIR compartmental model. Infections occur across edges on the static network layer at a constant rate $\beta_{s}$ whilst infections occur across edges on the dynamic network layer at a constant rate $\beta_{d}$. Infected individuals recover at a constant rate $\gamma$. Once recovered, a node cannot be reinfected, and can no longer transmit infection to its neighbours. A comprehensive list of model variables and parameters is given in Table \ref{tab:model_vars_params}. 

\begin{table}
\centering
\begin{tabular}{l | p{100mm}}
Variable/Parameter & Definition \\
\hline \hline $\beta_{s}$ & Per-edge disease transmission rate on static network layer \\
\hline $\beta_{d}$ & Per-edge disease transmission rate on dynamic network layer \\
\hline $\gamma$ & Per-individual disease recovery rate \\
\hline $\rho$ & The proportion of initially infectious individuals \\
\hline $\eta$ & Edge re-wiring rate on dynamic network layer \\
\hline $S(t),I(t),R(t)$ & The  susceptible, infectious and recovered proportion of the population at time $t\geq0$ \\
\hline $p_{s,t,d}$ & The proportion of individuals in the network that are a member of $s$ static line stubs,  $t$ triangle corners and $d$ dynamic line stubs \\
\hline $g(x,y,z)$ & Probability generating function for the numbers of static lines, triangles and dynamic lines of which an individual is a member \\
\hline $\theta_2(t)$ & A survivor function for remaining susceptible for some time $t\geq0$, given that the individual in question is a member of a single static line \\
\hline $\theta_3(t)$ & A survivor function for remaining susceptible for some time $t\geq0$, given that the individual in question is a member of a single triangle corner \\
\hline $\theta_{4}(t)$ & A survivor function for remaining susceptible for some time $t\geq0$, given that the individual in question is a member of a single dynamic line \\
\hline $\phi_{S}(t)$,$\phi_{I}(t)$,$\phi_{R}(t)$ & The probabilities that a neighbour of $u$ along a static line is susceptible, infectious or recovered, and has not transmitted infection to $u$  by time $t\geq0$ \\
\hline $\phi_{XY}(t)$ & The probability that two neighbours of $u$ in a triangle are in states $X$ and $Y \in \{S,I,R\}$ and have not transmitted infection to $u$ by time $t\geq0$ \\
\hline $A(t)$ & The rate (at time $t\geq0$) at which a random triangle neighbour $v$ of $u$ is infected from outside the triangle, given that $v$ was susceptible \\
\hline $B(t)$ & The rate (at time $t\geq0$) at which a random dynamic line stub neighbour $v$ of $u$ becomes infected from outside the dynamic line joining $u$ and $v$, given that $v$ was susceptible \\
\hline $\psi_{S}(t)$,$\psi_{I}(t)$,$\psi_{R}(t)$ & The probabilities (at time $t\geq0$) that a random dynamic stub belonging to $u$ has never been involved in transmitting infection to $u$ and is currently connected to a susceptible, infected, or recovered individual, respectively \\
\hline $\pi_{S}(t)$,$\pi_{I}(t)$,$\pi_{R}(t)$ & The probabilities (at time $t\geq0$) that a randomly chosen dynamic stub belongs to a susceptible, infected, or recovered individual, respectively \\
\end{tabular}
\caption{Definitions for model variables and parameters. Many definitions refer to the test node $u$, which is selected at random from the population and modified so that it cannot transmit infection, but can itself become infected.}
\label{tab:model_vars_params}
\end{table}

\subsection*{Edge-based compartmental model derivation}

We follow the edge-based compartmental modelling approach by considering the fate of a randomly selected test node $u$, which is prevented from transmitting infection. This assumption is a useful tool that eliminates conditional probability arguments that would need to be considered otherwise \cite{miller2012edge}. It does not introduce any approximation. At time zero, infection is introduced to a fraction $\rho$ of the population chosen uniformly at random, comprising the initial condition of the system. We assume that the test node $u$ is a member of $s$ static line stubs, $t$ static triangle corners and $d$ dynamic line stubs. Then the probability that $u$ is susceptible is $(1-\rho)\theta_{2}^{s}\theta_{3}^{t}\theta_{4}^{d}$, where $\theta_{2}$ is the probability that a random line (2-clique) on the static network layer has not transmitted infection to the test node, $\theta_{3}$ is the probability that neither of the other nodes in a random triangle on the static network layer have transmitted infection to the test node, and $\theta_{4}$ is the probability that a random \emph{stub} connected to $u$ on the dynamic network layer has never been involved in transmitting infection to the test node. Assuming we are able to calculate $\theta_{2}$, $\theta_{3}$ and $\theta_{4}$ as functions of time, we are able to calculate the proportion of susceptible individuals $S$ as a function of time. Given $S(t)$, we use $I(t)=1-S(t)-R(t)$ and $\dot{R}(t)=\gamma I(t)$ to calculate $I(t)$ and $R(t)$, completing the system.

\paragraph{Considering $\boldsymbol{\theta_{2}}$}

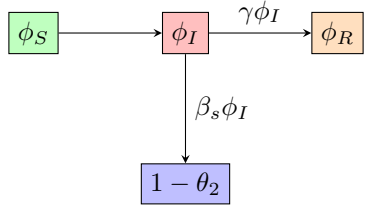
\begin{figure}
\centering
\begin{tikzpicture}[>=stealth]
\draw (0,0) node[draw=black, fill=green!25] (a) {$\phi_{S}$};
\draw (2,0) node[draw=black, fill=red!25] (b) {$\phi_{I}$};
\draw[->] (a) - - (b);
\draw (4,0) node[draw=black, fill=orange!25] (c) {$\phi_{R}$};
\draw[->] (b) - - node[auto] {$\gamma \phi_{I}$} (c);
\draw (2,-2) node[draw=black, fill=blue!25] (d) {$1-\theta_{2}$};
\draw[->] (b) - - node[auto] {$\beta_{s} \phi_{I}$} (d);
\end{tikzpicture}
\caption{{\bf Flow diagram for the flux of a static line partner through different states.}
The flux between the probabilities that the test node $u$ is connected by a line (2-clique) on the static network layer to a node $v$ that has not transmitted infection to $u$ and is susceptible ($\phi_{S}$), infectious ($\phi_{I}$) or recovered ($\phi_{R}$), and the probability that $v$ has transmitted infection to $u$, equal to $(1-\theta_2)$}
\label{fig:theta2compartments}
\end{figure}

We divide $\theta_{2}$ into $\phi_{S}$, $\phi_{I}$ and $\phi_{R}$, the probabilities that a random neighbour along a line on the static network layer has not transmitted infection to $u$, and is susceptible, infected, or recovered, respectively. The probability the neighbour has not transmitted infection to $u$ is $\theta_{2}=\phi_{S}+\phi_{I}+\phi_{R}$, and $(1-\theta_{2})$ is the probability that it has transmitted infection to $u$. The fluxes between these quantities is shown in Fig \ref{fig:theta2compartments}. The fluxes from $\phi_{I}$ to $\phi_{R}$ and from $\phi_{I}$ to $(1-\theta_{2})$ are proportional to one another. Both $\phi_{R}$ and $(1-\theta_{2})$ are equal to zero at time zero since we assume that no infection or recovery events can occur prior to time zero. By integrating the relation $\frac{d\phi_{R}}{dt}=\frac{\gamma}{\beta_{s}}\frac{d(1-\theta_{2})}{dt}$, and using the initial condition $\phi_{R}(0)=(1-\theta_{2}(0))=0$, we find the relation \begin{equation} \phi_{R}=\frac{\gamma}{\beta_{s}}(1-\theta_{2}). \label{eq:phi_R} \end{equation}

Next, we must calculate an expression for $\phi_{S}$. Consider the number of static line stubs attached to an individual that we reach by following a randomly chosen static line. Similarly, consider the number of static triangle corners attached to an individual reached by following a randomly chosen static triangle edge, and the number of dynamic line stubs attached to an individual we reach by following a randomly chosen dynamic line. Following edges in this way means we are more likely to arrive at an individual with a higher degree, in direct proportion to that individual's degree \cite{meyers2006predicting}. The random number of such lines and triangle corners is described by the \emph{excess degree distribution}, and we calculate the associated probability density functions for each edge type as follows. Denote $q_{s-1,t,d}\propto sp_{s,t,d}$ as the probability of there being $(s-1)$ static line stubs, $t$ triangle corners and $d$ dynamic line stubs connected to a susceptible node that we reach by following a static line, not counting the line by which we arrived. Similarly, denote $r_{s,t-1,d}\propto tp_{s,t,d}$ as the probability that if we follow a triangle edge to a susceptible node, there are $s$ static line stubs, $(t-1)$ triangle corners and $d$ dynamic line stubs connected to that node, not counting the triangle edge by which we arrived, and $w_{s,t,d-1}\propto dp_{s,t,d}$ as the probability that if we follow a dynamic edge to a susceptible node, there are $s$ static line stubs, $t$ triangle corners and $(d-1)$ dynamic line stubs connected to that node, not counting the dynamic edge by which we arrived. 

From above, we note that the probability that there are $s$ static line stubs, $t$ triangle corners and $d$ dynamic line stubs attached to a random neighbour of $u$ across a static line (not counting the line it was reached across) is $q_{s-1,t,d}\propto sp_{s,t,d}$. A neighbour reached by following a static line connected to $u$ is susceptible with probability $(1-\rho)\theta_{2}^{s-1}\theta_{3}^{t}\theta_{4}^{d}$ (recall that $u$ cannot transmit infection), where $s$, $t$ and $d$ are realisations of the excess degree distribution. We calculate $\phi_{S}$ by multiplying the probability that a random neighbour across a static line has $(s,t,d)$ neighbours, with the probability the random neighbour is susceptible, summing over all possible values of $(s,t,d)$, and dividing by $M=g^{(x)}(1,1,1)$, the expected number of static lines a randomly selected node belongs to. We find \begin{equation} \phi_{S} = \frac{(1-\rho)\sum_{s,t,d}s p_{s,t,d}\theta_{2}^{s-1}\theta_{3}^{t}\theta_{4}^{d}}{M}=\frac{(1-\rho)g^{(x)}(\theta_{2},\theta_{3},\theta_{4})}{g^{(x)}(1,1,1)}.	\label{eq:phi_S} \end{equation}	From the original definition of $\theta_{2}$ we have \begin{equation} \phi_{I}=\theta_{2}-\phi_{S}-\phi_{R}. \label{eq:phi_I} \end{equation} We are now able to calculate an expression for $\theta_{2}$ using equations \eqref{eq:phi_R}-\eqref{eq:phi_I}, and noting from Fig~\ref{fig:theta2compartments} that $\dot{\theta_{2}}=-\beta_{s} \phi_{I}$: \begin{equation} \dot{\theta_{2}}=-\beta_{s} \theta_{2} +\beta_{s} \frac{(1-\rho)g^{(x)}(\theta_{2},\theta_{3},\theta_{4})}{g^{(x)}(1,1,1)} +\gamma(1-\theta_{2}). \label{eq:d_theta_2} \end{equation}

\paragraph{Considering $\boldsymbol{\theta_{3}}$}

\begin{figure}
\centering
\begin{tikzpicture}[>=stealth]
\draw (0,0) node[draw=black, fill=green!25] (a) {$\phi_{SS}$};
\draw (3,0) node[draw=black, fill=red!25] (b) {$\phi_{SI}$};
\draw[->] (a) - - node[auto] {$2A\phi_{SS}$} (b);
\draw (6,1.5) node[draw=black, fill=orange!25] (c) {$\phi_{SR}$};
\draw[->] (b) - - node[auto] {$\gamma \phi_{SI}$} (c);
\draw (6,-1.5) node[draw=black, fill=red!25] (d) {$\phi_{II}$};
\draw[->] (b) - - node[auto,above] {$(A+\beta_{s})\phi_{SI}$} (d);
\draw (6,-5) node[draw=black, fill=blue!25] (e) {$1-\theta_{3}$};
\draw[->] (d) - - node[auto,above] {$2\beta_{s} \phi_{II}$} (e);
\draw[->] (b) - - node[auto,below] {$\beta_{s} \phi_{SI}$} (e);
\draw (9,0) node[draw=black, fill=red!25] (f) {$\phi_{IR}$};
\draw (12,0) node[draw=black, fill=orange!25] (g) {$\phi_{RR}$};
\draw[->] (c) - - node[auto,above] {$A\phi_{SR}$} (f);
\draw[->] (d) - - node[auto] {$2\gamma\phi_{II}$} (f);
\draw[->] (f) - - node[auto] {$\gamma\phi_{IR}$} (g);
\draw[->] (f) - - node[auto] {$\beta_{s}\phi_{IR}$} (e);
\end{tikzpicture}
\caption{{\bf Flow diagram for the flux of two triangle neighbours through different states.} The flux between the probabilities that the test node $u$ is connected in a triangle to two nodes in all possible disease status configurations, where neither triangle neighbour has transmitted infection to $u$, as well as the probability $(1-\theta_{3})$ that a node $v \neq u$ in the triangle has transmitted infection to the test node $u$}
\label{fig:theta3compartments}
\end{figure}
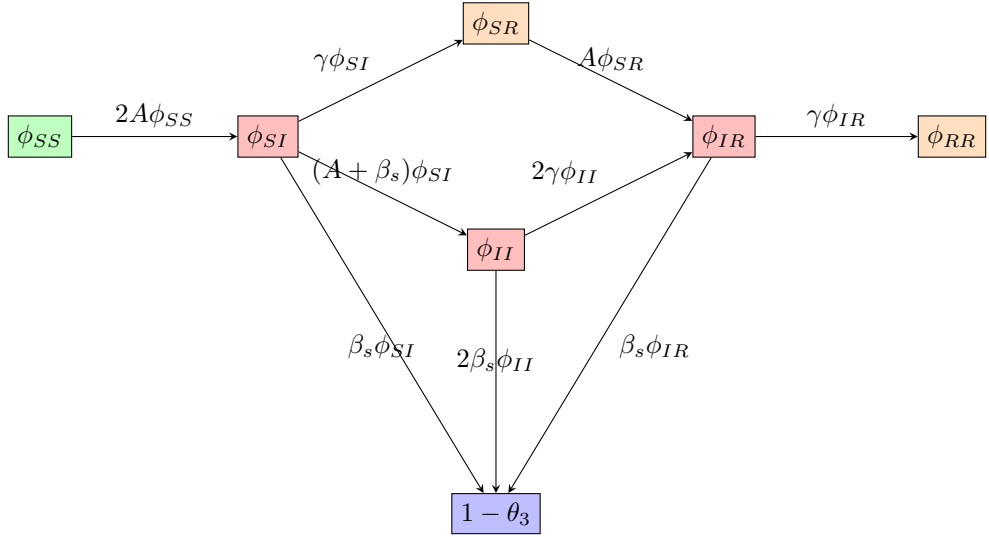

Since $\theta_{3}$ denotes the probability that neither of the other nodes in a triangle have transmitted infection to the test node, we must divide $\theta_{3}$ into six quantities $\phi_{SS}$, $\phi_{SI}$, $\phi_{SR}$, $\phi_{II}$, $\phi_{IR}$ and $\phi_{RR}$ in order to consider all possible disease status combinations for two individuals. For example, $\phi_{SI}$ denotes the probability that one triangle neighbour of $u$ is susceptible, whilst the other is infectious, and neither have transmitted infection to $u$. The flux between the various compartments can be seen in Fig~\ref{fig:theta3compartments}. There is no simple relation between $\phi_{RR}$ and $\theta_{3}$, so we take a different approach than before. We start with $\dot{\theta_{3}}$, which satisfies \begin{equation} \dot{\theta_{3}}=-\beta_{s} \phi_{SI}-2\beta_{s} \phi_{II}-\beta_{s} \phi_{IR}. \label{eq:d_theta_3} \end{equation}

To calculate elements in the right hand side of \eqref{eq:d_theta_3}, we must first obtain an expression for $\phi_{SS}$, the probability that both neighbours in a triangle are still susceptible. Under the assumption that no transmission events have occurred in the triangle, the probability that a \emph{single} triangle neighbour of $u$ is susceptible is \begin{equation*}(1-\rho)\sum_{s,t,d}tp_{s,t,d}\theta_{2}^{s}\theta_{3}^{t-1}\theta_{4}^{d}/\hat{M}=(1-\rho)g^{(y)}(\theta_{2},\theta_{3},\theta_{4})/g^{(y)}(1,1,1),\end{equation*} where $\hat{M}$ is the expected number of static triangle corners belonging to a randomly chosen individual. Since we require both triangle neighbours of $u$ to be susceptible, we have \begin{equation} \phi_{SS}=\left( \frac{(1-\rho)g^{(y)}(\theta_{2},\theta_{3},\theta_{4})}{g^{(y)}(1,1,1)} \right)^{2}. \label{eq:phi_SS} \end{equation} We choose $A$ to denote the rate at which a single triangle neighbour of $u$ becomes infected from \emph{outside} the triangle. From Fig~\ref{fig:theta3compartments} we know that $\frac{d\phi_{SS}}{dt}=-2A\phi_{SS}$, which implies $A=-\frac{d\phi_{SS}}{dt}/2\phi_{SS}$. To arrive at an explicit formula for $A$, we begin by calculating $\frac{d\phi_{SS}}{dt}$ via the chain rule: \begin{align*} \frac{d\phi_{SS}}{dt}&=2\left( \frac{(1-\rho)g^{(y)}(\theta_{2},\theta_{3},\theta_{4})}{g^{(y)}(1,1,1)} \right) \frac{d}{dt}\left( \frac{(1-\rho)g^{(y)}(\theta_{2},\theta_{3},\theta_{4})}{g^{(y)}(1,1,1)}  \right) \\ &=\frac{2(1-\rho)^{2}g^{(y)}(\theta_{2},\theta_{3},\theta_{4})}{g^{(y)}(1,1,1)}\left( \frac{g^{(y)}(1,1,1)\left( g^{(y)}(\theta_{2},\theta_{3},\theta_{4}) \right)'-g^{(y)}(\theta_{2},\theta_{3},\theta_{4})\left( g^{(y)}(1,1,1) \right)'}{\left( g^{(y)}(1,1,1) \right)^{2}} \right). \end{align*} We know that $\left( g^{(y)}(1,1,1) \right)'=0$, since $g^{(y)}(1,1,1)=\sum_{s,t,d}tp_{s,t,d} \in \mathbb{R}$. Hence \begin{align*} \frac{d\phi_{SS}}{dt}&=\frac{2(1-\rho)^{2}g^{(y)}(\theta_{2},\theta_{3},\theta_{4})}{g^{(y)}(1,1,1)}\left( \frac{g^{(y)}(1,1,1)\left( g^{(y)}(\theta_{2},\theta_{3},\theta_{4}) \right)'}{\left( g^{(y)}(1,1,1) \right)^{2}} \right) \\ &=\frac{2(1-\rho)^{2}g^{(y)}(\theta_{2},\theta_{3},\theta_{4})}{g^{(y)}(1,1,1)}\left( \frac{\left( g^{(y)}(\theta_{2},\theta_{3},\theta_{4}) \right)'}{g^{(y)}(1,1,1)} \right). \end{align*} Next, we calculate $\left( g^{(y)}(\theta_{2},\theta_{3},\theta_{4}) \right)'$ using  $\frac{dg(x,y,z)}{dt}=\frac{\partial g}{\partial x}\frac{dx}{dt}+\frac{\partial g}{\partial y}\frac{dy}{dt}+\frac{\partial g}{\partial z}\frac{dz}{dt}$ to obtain \[\left( g^{(y)}(\theta_{2},\theta_{3},\theta_{4}) \right)'= g^{(y,x)}(\theta_{2},\theta_{3},\theta_{4})\dot{\theta_{2}}+g^{(y,y)}(\theta_{2},\theta_{3},\theta_{4})\dot{\theta_{3}}+g^{(y,z)}(\theta_{2},\theta_{3},\theta_{4})\dot{\theta_{4}}.\] Thus we have \[ \frac{d\phi_{SS}}{dt}= \frac{2(1-\rho)^{2}g^{(y)}(\theta_2,\theta_3,\theta_4)}{g^{(y)}(1,1,1)}\left( \frac{g^{(y,x)}(\theta_{2},\theta_{3},\theta_{4})\dot{\theta_{2}}+g^{(y,y)}(\theta_{2},\theta_{3},\theta_{4})\dot{\theta_{3}}+g^{(y,z)}(\theta_{2},\theta_{3},\theta_{4})\dot{\theta_{4}}}{g^{(y)}(1,1,1)} \right).\] Using $A=-\frac{d\phi_{SS}}{dt}/2\phi_{SS}$ and some simplification, we find an explicit formula for $A$: \begin{equation} A=-\left(\frac{g^{(y,x)}(\theta_2,\theta_3,\theta_4)\dot{\theta_2}+g^{(y,y)}(\theta_2,\theta_3,\theta_4)\dot{\theta_3}+g^{(y,z)}(\theta_2,\theta_3,\theta_4)\dot{\theta_4}}{g^{(y)}(\theta_2,\theta_3,\theta_4)}\right). \label{eq:A} \end{equation}

Now we are ready to calculate equations for $\phi_{SI}$, $\phi_{II}$ and $\phi_{IR}$. We also require $\phi_{SR}$, but do not require a formula for $\phi_{RR}$. Using the flow diagram in Fig~\ref{fig:theta3compartments}, we have \begin{align} \dot{\phi_{SI}}&=2A\phi_{SS}-(A+2\beta_{s}+\gamma)\phi_{SI}, \label{eq:d_phi_SI} \\ \dot{\phi_{SR}}&=\gamma \phi_{SI}-A\phi_{SR}, \label{eq:d_phi_SR} \\ \dot{\phi_{II}}&=(A+\beta_{s})\phi_{SI}-2(\beta_{s}+\gamma)\phi_{II}, \label{eq:d_phi_II} \\ \dot{\phi_{IR}}&=A\phi_{SR}+2\gamma \phi_{II}-(\beta_{s}+\gamma)\phi_{IR}. \label{eq:d_phi_IR} \end{align}

\paragraph{Considering $\boldsymbol{\theta_{4}}$}

\begin{figure}
\centering
\begin{tikzpicture}[>=stealth]
\draw (0,0) node[draw=black, fill=blue!25] (a) {$1-\theta_{4}$};
\draw (0,3) node[draw=black, fill=red!25] (b) {$\psi_{I}$};
\draw (-3,3) node[draw=black, fill=green!25] (c) {$\psi_{S}$};
\draw (3,3) node[draw=black, fill=orange!25] (d) {$\psi_{R}$};
\draw (0,6) node[draw=black, rounded corners, fill=white!25] (e) {$\eta \theta_{4}$};
\draw[->] (b) - - node[auto,left] {$\beta_{d} \psi_{I}$} (a);
\draw[->] (b) - - node[auto,below] {$\gamma \psi_{I}$} (d);
\draw[->] (c) - - node[auto,below] {$B\psi_{S}$} (b);
\draw [->, rounded corners] (d) -- node[auto,right] {$\eta \psi_{R}$} (3,6) -- (e);
\draw[->] (e) - - node[auto,above] {$\eta \theta_{4}\pi_{R}$} (d);
\draw[->,rounded corners] (c) -- node[auto,left] {$\eta \psi_{S}$} (-3,6) -- (e);
\draw[->] (e) - - node[auto,above] {$\eta \theta_{4}\pi_{S}$} (c);
\draw[->, rounded corners] (b) -- node[auto,right] {$\eta \psi_{I}$} (0.5,4.5) -- (e);
\draw[->, rounded corners] (e) -- (-0.5,4.5) -- node[auto,left] {$\eta \theta_{4} \pi_{I}$} (b); 
\end{tikzpicture}
\caption{{\bf Flow diagram for the flux of a dynamic edge partner through different states.} The flux between the probabilities $\theta_{4}=\psi_{S}+\psi_{I}+\psi_{R}$ that a random stub currently connected to $u$ on the dynamic network layer has never been involved in transmitting infection to $u$. Note that the compartment denoted $\eta \theta_{4}$ is not a compartment in the typical sense. When edges break (at rate $\eta$) in the model, moving into `compartment' $\eta \theta_{4}$, new edges are formed \emph{immediately} without delay, moving straight back into compartments $\psi_{S}$, $\psi_{I}$ or $\psi_{R}$. $\pi_{S}$, $\pi_{I}$ and $\pi_{R}$ denote the probabilities that a randomly chosen dynamic stub belongs to a susceptible, infected, or recovered node, respectively}
\label{fig:theta4compartments}
\end{figure}
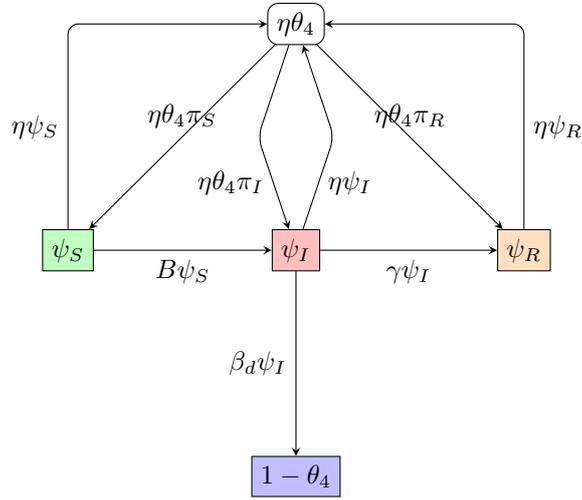

To take into account the dynamic rewiring of edges, we introduce $\theta_{4}=\psi_{S}+\psi_{I}+\psi_{R}$, where $\psi_{I}$ denotes the probability that a random dynamic stub belonging to the test node $u$ has never been involved in transmitting infection to $u$, and is \emph{currently connected} to an infectious node. Other important assumptions with respect to dynamic edge rewiring are the following: we assume that when one partnership ends, a new partnership forms immediately, neglecting any between-partner period, and we assume that edges break at rate $\eta$. The flux between the various compartments of interest can be seen in Fig~\ref{fig:theta4compartments}.

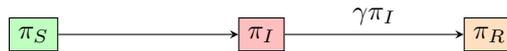
\begin{figure}
\centering
\begin{tikzpicture}[>=stealth]
\draw (0,0) node[draw=black,fill=green!25] (a) {$\pi_{S}$};
\draw (3,0) node[draw=black,fill=red!25] (b) {$\pi_{I}$};
\draw (6,0) node[draw=black,fill=orange!25] (c) {$\pi_{R}$};
\draw[->] (a) - - (b);
\draw[->] (b) - - node[auto,above] {$\gamma \pi_{I}$} (c);
\end{tikzpicture}
\caption{{\bf Flow diagram for the flux of a dynamic line stub through different states.} The flux between $\pi_{S}$, $\pi_{I}$ and $\pi_{R}$, the probabilities that a randomly chosen dynamic stub belongs to a susceptible, infected or recovered node, respectively} \label{fig:picompartments}
\end{figure}

Previously, $\phi_{S}$ (which corresponds to $\psi_{S}$ in this subsection) was calculated explicitly as the probability that the neighbour is susceptible. With dynamic edge re-wiring, an edge that previously transmitted infection may later become connected to a susceptible node, so the previous calculation of $\phi_{S}$ does not apply here. To find $\psi_{S}$, we need to calculate the probability that a \emph{newly formed} edge connects to a susceptible, infectious, or recovered individual. We call these probabilities $\pi_{S}$, $\pi_{I}$ and $\pi_{R}$ and note that they are equivalent to the probabilities that a randomly chosen dynamic stub belongs to a node in each disease compartment. The flux between these probabilities can be seen in Fig~\ref{fig:picompartments}.

First, we calculate the values $\pi_{S}$, $\pi_{I}$ and $\pi_{R}$, beginning with $\pi_{S}$. If we select a dynamic stub at random, the probability that it belongs to an individual partaking in $s$ static lines, $t$ triangles and $d$ dynamic stubs is $dp_{s,t,d}/\tilde{M}$, where $\tilde{M}=g^{(z)}(1,1,1)$ is the expected number of dynamic edges that a random individual belongs to. At time zero, infection is introduced at random to a proportion $\rho$ of the population. Thus the probability of any node being susceptible at time zero is $(1-\rho)$. The probability of a node with degree $(s,t,d)$ being susceptible after some time, given that it was susceptible at time zero, is $\theta_{2}^{s}\theta_{3}^{t}\theta_{4}^{d}$. Hence $\pi_{S}=(1-\rho)\sum_{s,t,d}p_{s,t,d}d\theta_{2}^{s}\theta_{3}^{t}\theta_{4}^{d}/\tilde{M}$, with the summation taken over all degree possibilities described by the probability mass function $p_{s,t,d}$. Stubs belonging to infected nodes become stubs belonging to recovered nodes at rate $\gamma$, hence $\dot{\pi_{R}}=\gamma \pi_{I}$, and $\pi_{I}=1-\pi_{S}-\pi_{R}$. The equation for $\pi_{S}$ can be condensed using the PGF \eqref{eq:pgf_g}, so we have \begin{align}	\pi_{S}&=\frac{(1-\rho)\sum_{s,t,d}p_{s,t,d}d\theta_{2}^{s}\theta_{3}^{t}\theta_{4}^{d}}{\sum_{s,t,d}p_{s,t,d}d}=\frac{(1-\rho)\theta_{4}g^{(z)}(\theta_{2},\theta_{3},\theta_{4})}{g^{(z)}(1,1,1)}, \label{eq:pi_S} \\	\pi_{I}&= 1 - \pi_{S}  - \pi_{R}, \label{eq:pi_I} \\ \dot{\pi_{R}}&=\gamma \pi_{I}. \label{eq:pi_R}	\end{align}

To complete the system we need to calculate the flux $B\psi_{S}$ from $\psi_{S}$ to $\psi_{I}$ by solving a differential equation for $\psi_{S}$. $B$ describes the rate at which a susceptible dynamic-edge neighbour $v$ of $u$ becomes infected from outside the dynamic edge joining $u$ and $v$. Consider a random test node $u$ and a random dynamic-edge neighbour $v$ of $u$, at some time $t$. Let $\zeta$ denote the probability that the two stubs joining $u$ and $v$ have not previously been involved in transmitting infection to $u$ or to $v$, prior to the $u-v$ edge forming. The probability that $v$ is susceptible and that $u$'s stub has not previously transmitted to $u$ is $\zeta(1-\rho)\theta_{2}^{s}\theta_{3}^{t}\theta_{4}^{d-1}$, where $s$ is the number of static lines $v$ partakes in, $t$ is the number of triangles $v$ partakes in, and $d$ is the dynamic line stub degree of $v$. Since we do not know the values $(s,t,d)$ for $v$, we must consider all possible combinations of degrees. The probability of a randomly chosen dynamic stub belonging to a node with degree $(s,t,d)$ is $dp_{s,t,d}/g^{(z)}(1,1,1)$. We conclude that \[ \psi_{S}=\frac{\zeta(1-\rho)\sum_{s,t,d}p_{s,t,d}d\theta_{2}^{s}\theta_{3}^{t}\theta_{4}^{d-1}}{g^{(z)}(1,1,1)}=\frac{\zeta(1-\rho)g^{(z)}(\theta_{2},\theta_{3},\theta_{4})}{g^{(z)}(1,1,1)}. \] 

To calculate the derivative of $\psi_{S}$, we first consider the derivative of $\zeta$. This is given by subtracting the rate at which such edges break, $\eta \zeta$, from the rate at which such edges form, $\eta \theta_{4}^{2}$ (one $\theta_{4}$ for $u$'s stub and one for $v$'s stub). We have \[ \dot{\zeta}=\eta \theta_{4}^{2} -\eta \zeta. \] We have an expression for $\dot{\zeta}$, so the derivative of $\psi_{S}$ can be found via the chain rule: \begin{align*} \dot{\psi_{S}}&=\dot{\zeta} \frac{(1-\rho)g^{(z)}(\theta_{2},\theta_{3},\theta_{4})}{g^{(z)}(1,1,1)}+\frac{\zeta(1-\rho)}{g^{(z)}(1,1,1)}\left( g^{(z)}(\theta_{2},\theta_{3},\theta_{4}) \right)' \\ &=\frac{\eta \theta_{4}^{2}(1-\rho)g^{(z)}(\theta_{2},\theta_{3},\theta_{4})}{g^{(z)}(1,1,1)}-\frac{\eta \zeta(1-\rho)g^{(z)}(\theta_{2},\theta_{3},\theta_{4})}{g^{(z)}(1,1,1)}+\frac{\zeta(1-\rho)}{g^{(z)}(1,1,1)}\left( g^{(z)}(\theta_{2},\theta_{3},\theta_{4}) \right)' \\ &=\pi_{S}\eta\theta_{4}-\eta\psi_{S}+\frac{\zeta(1-\rho)}{g^{(z)}(1,1,1)}\left( g^{(z)}(\theta_{2},\theta_{3},\theta_{4}) \right)'  \\ &=\pi_{S}\eta\theta_{4}-\eta\psi_{S}+\frac{\zeta(1-\rho)}{g^{(z)}(1,1,1)}\left( g^{(z,x)}(\theta_{2},\theta_{3},\theta_{4})\dot{\theta_{2}}+g^{(z,y)}(\theta_{2},\theta_{3},\theta_{4})\dot{\theta_{3}}+g^{(z,z)}(\theta_{2},\theta_{3},\theta_{4})\dot{\theta_{4}} \right) \\ &=\pi_{S}\eta\theta_{4}-\eta\psi_{S}+\frac{\psi_{S}\left( g^{(z,x)}(\theta_{2},\theta_{3},\theta_{4})\dot{\theta_{2}}+g^{(z,y)}(\theta_{2},\theta_{3},\theta_{4})\dot{\theta_{3}}+g^{(z,z)}(\theta_{2},\theta_{3},\theta_{4})\dot{\theta_{4}} \right)}{g^{(z)}(\theta_{2},\theta_{3},\theta_{4})}, \end{align*} with simplifications achieved by utilising $\pi_{S}=(1-\rho)\theta_{4}g^{(z)}(\theta_{2},\theta_{3},\theta_{4})/g^{(z)}(1,1,1)$ and $\psi_{S}=\zeta(1-\rho)g^{(z)}(\theta_{2},\theta_{3},\theta_{4})/g^{(z)}(1,1,1)$. From Fig~\ref{fig:theta4compartments} we have $\dot{\psi_{S}}=\eta \theta_{4}\pi_{S}-\eta\psi_{S}-B\psi_{S}$, so we calculate the flux between compartments $\psi_{S}$ and $\psi_{I}$ using the rate \begin{equation}B=-\left(\frac{ g^{(z,x)}(\theta_{2},\theta_{3},\theta_{4})\dot{\theta_{2}}+g^{(z,y)}(\theta_{2},\theta_{3},\theta_{4})\dot{\theta_{3}}+g^{(z,z)}(\theta_{2},\theta_{3},\theta_{4})\dot{\theta_{4}}}{g^{(z)}(\theta_{2},\theta_{3},\theta_{4})}\right). \label{eq:B} \end{equation}

The $\psi_{S}$ to $\psi_{I}$ flux is the product of $\psi_{S}$, the probability that a random dynamic stub has not transmitted infection to the test node $u$ and is currently connected to a susceptible node, with rate $B$, the rate that a neighbouring susceptible node $v$ becomes infected from outside the dynamic edge, given that the stub has not transmitted and connects $u$ to a susceptible node. Following the flow diagram in Fig~\ref{fig:theta4compartments}, we have the differential equations \begin{align} \dot{\theta_{4}}&= -\beta_{d} \psi_{I}, \label{eq:d_theta_4} \\ \dot{\psi_{S}}&= \eta \theta_{4} \pi_{S} - (B+\eta)\psi_{S}, \label{eq:d_psi_S} \\ \dot{\psi_{I}}&= B \psi_{S}+\eta \theta_{4}\pi_{I}-(\eta+\gamma+\beta_{d})\psi_{I}, \label{eq:d_psi_I} \\ \dot{\psi_{R}}&= \gamma \psi_{I}+\eta \theta_{4}\pi_{R}-\eta\psi_{R}. \label{eq:d_psi_R} \end{align}

\paragraph{Population-level equations}

We began the EBCM derivation by considering the probability of a randomly selected test node $u$ (which is prevented from transmitting infection) being susceptible as $\theta_{2}^{s}\theta_{3}^{t}\theta_{4}^{d}$, given that the node has degree $(s,t,d)$. Since we have calculated formulae for $\theta_{2}$, $\theta_{3}$ and $\theta_{4}$, we can derive population-level equations describing the proportion of the population in each disease compartment at each point in time: \begin{align}	S(t) &= (1-\rho)g(\theta_{2}(t),\theta_{3}(t),\theta_{4}(t))=(1-\rho)\sum_{s,t,d}p_{s,t,d}\theta_{2}(t)^{s}\theta_{3}(t)^{t}\theta_{4}(t)^{d}, \label{eq:S} \\ I(t) &= 1 - S(t) - R(t), \label{eq:I} \\	\dot{R}(t) &= \gamma I(t). \label{eq:R}	\end{align} Equations \eqref{eq:pgf_g}-\eqref{eq:R} form a complete system describing an SIR epidemic spreading across a dual-layer multiplex network consisting of a static network layer constructed from line stubs and triangle corners and a dynamic network layer constructed from line stubs only, where edges rewire and degrees are conserved.

\subsection*{Deriving the basic reproduction number $\boldsymbol{R_{0}}$}

The basic reproduction number $R_{0}$ is defined as the average number of infections caused by a single infectious individual, early in an epidemic process, in an otherwise susceptible population. In the model, a multiplex network structure is generated using three distinct edge distributions (static line stubs, static triangle corners and dynamic line stubs). To compute $R_{0}$ we must consider the average number of infections caused across each type of edge, whilst also considering the type of edge that the infection was originally \emph{received} across. With 3 edge types, this constitutes 9 values, grouped together to form the next generation matrix \[ \boldsymbol{G}= \left( \begin{array}{c c c} G_{ss} & G_{st} & G_{sd} \\ G_{ts} & G_{tt} & G_{td} \\ G_{ds} & G_{dt} & G_{dd} \end{array}\right), \] where matrix element $G_{ij}$ describes the average number of infections caused across edges of type $j$, where the infector \emph{received} infection across an edge of type $i$. Following the next generation matrix approach \cite{diekmann2009construction}, the value of $R_{0}$ is found via the leading eigenvalue of the matrix $\boldsymbol{G}$, or equivalently, the eigenvalue with greatest magnitude.

To find $R_{0}$, we begin by deriving expressions for values in the first column of $\boldsymbol{G}$. Firstly, consider the non-diagonal matrix entries $G_{ts}$ and $G_{ds}$. We want to compute the expected number of infection events occurring across static lines, when individuals contracted infection across a triangle edge or a dynamic line. In both cases, we require the expected static line stub degree, multiplied by the expected number of infections caused across a \emph{single} static line attached to the infectious individual. Say the expected static line stub degree is denoted $\langle k_{s} \rangle$. Now we require the expected number of infections caused across a single static edge attached to an infectious individual, in an otherwise susceptible population. A single static edge joining a susceptible and an infectious individual, in an otherwise susceptible population, has two event possibilities: a single recovery, or a single infection. Denote $X$ as the random variable describing the number of infection events occurring across a single static line joining a susceptible to an infectious individual, in an otherwise susceptible population. Using the expectation formula, and since there can only be zero or one infection events occurring across such an edge, we find the expected number of infections across a static line joining a susceptible to an infectious individual simply as \begin{equation*} 0\cdot\mathbb{P}(X=0)+1\cdot\mathbb{P}(X=1) = \mathbb{P}(X=1). \end{equation*} The probability of a single infection occurring across such a static edge, prior to any recovery, is $\frac{\beta_{s}}{\beta_{s}+\gamma}$. Thus we can say that $G_{ts}=G_{ds}=\langle k_{s}\rangle \frac{\beta_{s}}{\beta_{s}+\gamma}$. 

Finally, we calculate an expression for the diagonal matrix element $G_{ss}$, by multiplying the expected \emph{excess} static line stub degree, denoted $\langle s\rangle$, by the expected number of infections caused across a single static line joining a susceptible individual to an infectious individual in an otherwise susceptible population. Following the same argument for $G_{ts}$ and $G_{ds}$, we compute the expected number of infection events for $G_{ss}$ as $\frac{\beta_{s}}{\beta_{s}+\gamma}$, and we obtain $G_{ss}=\langle s\rangle \frac{\beta_{s}}{\beta_{s}+\gamma}$. 

Next we derive expressions for the values $G_{st}$, $G_{tt}$ and $G_{dt}$ in the second column of the matrix $\boldsymbol{G}$. We firstly consider the non-diagonal elements $G_{st}$ and $G_{dt}$. Both $G_{st}$ and $G_{dt}$ are calculated by multiplying the expected triangle \emph{corner} degree, denoted $\langle k_{t}\rangle$, by the expected number of infection events caused within a \emph{single} triangle attached to an infectious node in an otherwise susceptible population. In a single triangle comprised of two susceptible individuals attached to an infectious individual, there are a finite number of infection event possibilities: either no further infections occur (the infectious individual recovers), one infection event occurs, or two infection events occur. Define $Y$ as the random variable describing the number of infection events within such a triangle. Using the expectation formula, we find the expected number of infection events within a triangle comprised of two susceptible individuals and an infective, in an otherwise susceptible population, as \begin{equation*} 0\cdot\mathbb{P}(Y=0)+1\cdot\mathbb{P}(Y=1)+2\cdot\mathbb{P}(Y=2)=\mathbb{P}(Y=1)+2\cdot\mathbb{P}(Y=2). \end{equation*} To continue, we must compute the probabilities $\mathbb{P}(Y=1)$ and $\mathbb{P}(Y=2)$ explicitly. $\mathbb{P}(Y=1)$ describes the probability that the original infective infects \emph{one} out of two triangle neighbours. In this case, either one of the two susceptible neighbours can become infectious, and both infectious triangle members must then recover, so that it is impossible for any more than one infection event to occur. In a triangle comprised of a single infective and two susceptible nodes, there are four distinct nodal orders in which a single infection event is followed by the recovery of both infectious nodes. We find \begin{eqnarray*} \mathbb{P}(Y=1)&=&\frac{4\beta_{s}}{2\beta_{s}+\gamma}\left(\frac{\gamma}{2\beta_{s}+2\gamma}\right)\left(\frac{\gamma}{\beta_{s}+\gamma}\right) \\ &=&\frac{2\beta_{s}}{2\beta_{s}+\gamma}\left(\frac{2\gamma}{2\beta_{s}+2\gamma}\right)\left(\frac{\gamma}{\beta_{s}+\gamma}\right) \\ &=&\frac{2\beta_{s}}{2\beta_{s}+\gamma}\left(\frac{\gamma}{\beta_{s}+\gamma}\right)\left(\frac{\gamma}{\beta_{s}+\gamma}\right) \\ &=&\frac{2\beta_{s}}{2\beta_{s}+\gamma}\left(\frac{\gamma}{\beta_{s}+\gamma}\right)^{2}. \end{eqnarray*} 

Considering $\mathbb{P}(Y=2)$ is more complex, as there are two distinct ways in which two infection events can occur in a triangle between an infective and two susceptible individuals. Firstly, the original infective can infect both of its triangle neighbours consecutively, prior to any recovery events. The probability of both triangle infection events occurring in succession is given by $\left(\frac{2\beta_{s}}{2\beta_{s}+\gamma}\right) \left(\frac{2\beta_{s}}{2\beta_{s}+2\gamma}\right)=\left(\frac{2\beta_{s}}{2\beta_{s}+\gamma}\right)\left(\frac{\beta_{s}}{\beta_{s}+\gamma}\right)$. Secondly, the original infective can cause two triangle infections via \emph{three} consecutive events. In this case, the originally infectious triangle member firstly infects one susceptible triangle neighbour at rate $\frac{2\beta_{s}}{2\beta_{s}+\gamma}$. The triangle is now comprised of two infectious individuals attached to a single susceptible individual. The second event to occur is a recovery of either the original infector or its first infectee, occurring at rate $\frac{2\gamma}{2\beta_{s}+2\gamma}=\frac{\gamma}{\beta_{s}+\gamma}$. The triangle is now comprised of a susceptible, an infective, and a recovered individual, in an otherwise susceptible population. Following the recovery event, the final event is an infection of the remaining susceptible triangle member, occurring at rate $\frac{\beta_{s}}{\beta_{s}+\gamma}$. The probability of all three events occurring in succession is thus $\frac{2\beta_{s}}{2\beta_{s}+\gamma}\left(\frac{\gamma}{\beta_{s}+\gamma}\right)\left(\frac{\beta_{s}}{\beta_{s}+\gamma}\right)$.

In the latter case of an infection, followed by a recovery, followed by another infection within a triangle originally composed of an infective and two susceptible individuals in an otherwise susceptible population, the original infector may not be directly involved in every single infection event. However, for the purposes of deriving $R_{0}$, we say that the original infector \emph{caused} these infections, regardless of the order in which triangle members recover and infect one another.

Since there are two distinct ways in which two infections can take place within a triangle comprised of an infective and two susceptible individuals, we take the sum of both individual probabilities to obtain $\mathbb{P}(Y=2)$: \begin{eqnarray*} \mathbb{P}(Y=2) &=& \frac{2\beta_{s}}{2\beta_{s}+\gamma}\left(\frac{\beta_{s}}{\beta_{s}+\gamma}\right)+\frac{2\beta_{s}}{2\beta_{s}+\gamma}\left(\frac{\gamma}{\beta_{s}+\gamma}\right)\left(\frac{\beta_{s}}{\beta_{s}+\gamma}\right) \\ &=& \frac{2\beta_{s}}{2\beta_{s}+\gamma}\left(\frac{\beta_{s}}{\beta_{s}+\gamma}\right)\left[1+\frac{\gamma}{\beta_{s}+\gamma}\right]. \end{eqnarray*} We find the expected number of infection events within a triangle comprised of two susceptible individuals and an infective, in an otherwise susceptible population, as \begin{eqnarray*} \mathbb{P}(Y=1)+2\cdot\mathbb{P}(Y=2) &=& \frac{2\beta_{s}}{2\beta_{s}+\gamma}\left(\frac{\gamma}{\beta_{s}+\gamma}\right)^{2} + \frac{4\beta_{s}}{2\beta_{s}+\gamma}\left(\frac{\beta_{s}}{\beta_{s}+\gamma}\right)\left[1+\frac{\gamma}{\beta_{s}+\gamma}\right] \\ &=& \frac{2\beta_{s}}{2\beta_{s}+\gamma}\left[ \left(\frac{\gamma}{\beta_{s}+\gamma}\right)^{2}+2\left(\frac{\beta_{s}}{\beta_{s}+\gamma}\right)\left[1+\frac{\gamma}{\beta_{s}+\gamma}\right] \right] \\ &=& \frac{2\beta_{s}}{2\beta_{s}+\gamma}\left[\left(\frac{\gamma}{\beta_{s}+\gamma}\right)^{2}+2\left[1-\frac{\gamma}{\beta_{s}+\gamma}\right]\left[1+\frac{\gamma}{\beta_{s}+\gamma}\right]\right] \\ &=& \frac{2\beta_{s}}{2\beta_{s}+\gamma}\left[\left(\frac{\gamma}{\beta_{s}+\gamma}\right)^{2}+2\left[1-\left(\frac{\gamma}{\beta_{s}+\gamma}\right)^{2}\right]\right] \\ &=& \frac{2\beta_{s}}{2\beta_{s}+\gamma}\left[2-\left(\frac{\gamma}{\beta_{s}+\gamma}\right)^{2}\right]. \end{eqnarray*} Then we have $G_{st}=\langle k_{t}\rangle \frac{2\beta_{s}}{2\beta_{s}+\gamma}\left[2-\left(\frac{\gamma}{\beta_{s}+\gamma}\right)^{2}\right]=G_{dt}$, where $\langle k_{t}\rangle$ denotes the expected static triangle \emph{corner} degree. Finally, we have $G_{tt}=\langle t\rangle \frac{2\beta_{s}}{2\beta_{s}+\gamma}\left[2-\left(\frac{\gamma}{\beta_{s}+\gamma}\right)^{2}\right]$, where $\langle t\rangle$ denotes the expected \emph{excess} static triangle corner degree.

We conclude by deriving elements from the third column of $\boldsymbol{G}$, starting with non-diagonal matrix elements $G_{sd}$ and $G_{td}$. In both cases, we multiply the expected dynamic line stub degree, denoted $\langle k_{d}\rangle$, by the expected number of infection events occurring across a single dynamic line stub attached to an infectious individual, in an otherwise susceptible population. 

The probability of a dynamic stub attached to an infective in an otherwise susceptible population transmitting infection \emph{at least once} is $\frac{\beta_{d}}{\beta_{d}+\gamma}$. If such an infection occurs, the I-S pairing becomes an I-I pairing with a dynamic edge joining the two individuals. The probability of a dynamic I-I edge rewiring, prior to any recovery event, is $\frac{\eta}{\eta+\gamma}$. We can assume that any I-I edge rewires to become an I-S edge in the limit of large population size, since we are early on in an epidemic process, and we began with an otherwise susceptible population. The probability that an infectious dynamic stub infects its new susceptible neighbour is $\frac{\beta_{d}}{\beta_{d}+\gamma}$. This rewiring and infecting process can occur an arbitrary number of times in the model. The expected number of infections of this type can be calculated by taking the sum \[ \sum_{n=0}^{\infty}\frac{\beta_{d}}{\beta_{d}+\gamma}r^{n} = \frac{\beta_{d}}{\beta_{d}+\gamma}\left( \frac{1}{1-r}\right), \] by the geometric series, and where $r$ is defined as $\frac{\eta\beta_{d}}{(\eta+\gamma)(\beta_{d}+\gamma)}$, the probability of an infectious individual's dynamic edge rewiring, followed immediately by its dynamic stub infecting the new (susceptible) neighbour across the rewired edge. We obtain the matrix values $G_{sd}=\langle k_{d}\rangle \frac{\beta_{d}}{\beta_{d}+\gamma}\left( \frac{1}{1-r}\right)=G_{td}$.

Finally, we compute $G_{dd}$, defined as the expected number of infections caused across dynamic edges, where the infector received infection across a dynamic edge itself. Firstly, consider the \emph{single} dynamic I-I edge which originally infected our individual. The probability of the edge rewiring, leaving our infective in an I-S dynamic edge pairing, is $\frac{\eta}{\eta+\gamma}$. The probability of the infectious dynamic stub infecting the new susceptible neighbour is $\frac{\beta_{d}}{\beta_{d}+\gamma}$. Thus the probability that the dynamic stub which originally contracted infection infects $\geq n$ individuals is $r^{n}$, where $r=\frac{\eta\beta_{d}}{(\eta+\gamma)(\beta_{d}+\gamma)}$. We compute the expected number of infections of this type by taking the sum of $r^{n}$ for $n=1:\infty$ \[ \sum_{n=1}^{\infty}r^{n} = \frac{r}{1-r}, \] by the geometric series. Now consider the remaining dynamic edges associated with our infectious individual. We require the expected number of infections caused by a single edge of this type. Using the same argument as for $G_{sd}$ and $G_{td}$, we find the expected number of infections caused by one dynamic edge attached to our infectious individual as $\frac{\beta_{d}}{\beta_{d}+\gamma}\left(\frac{1}{1-r}\right)$. Thus we find $G_{dd}=\frac{r}{1-r}+\langle d\rangle\frac{\beta_{d}}{\beta_{d}+\gamma}\left(\frac{1}{1-r}\right)$, where $\langle d\rangle$ is the expected excess dynamic line stub degree. 

In detail, the next generation matrix $\boldsymbol{G}$ takes the form \begin{equation} \left( \begin{array}{c c c} \langle s\rangle \frac{\beta_{s}}{\beta_{s}+\gamma} & \langle k_{t}\rangle \frac{2\beta_{s}}{2\beta_{s}+\gamma}\left(2-\left(\frac{\gamma}{\beta_{s}+\gamma}\right)^{2}\right) & \langle k_{d}\rangle \frac{\beta_{d}}{\beta_{d}+\gamma}\left( \frac{1}{1-r}\right) \\ \langle k_{s}\rangle \frac{\beta_{s}}{\beta_{s}+\gamma} & \langle t\rangle \frac{2\beta_{s}}{2\beta_{s}+\gamma}\left(2-\left(\frac{\gamma}{\beta_{s}+\gamma}\right)^{2}\right) & \langle k_{d}\rangle \frac{\beta_{d}}{\beta_{d}+\gamma}\left( \frac{1}{1-r}\right) \\ \langle k_{s}\rangle \frac{\beta_{s}}{\beta_{s}+\gamma} & \langle k_{t}\rangle \frac{2\beta_{s}}{2\beta_{s}+\gamma}\left(2-\left(\frac{\gamma}{\beta_{s}+\gamma}\right)^{2}\right) & \frac{r}{1-r}+\langle d\rangle\frac{\beta_{d}}{\beta_{d}+\gamma}\left(\frac{1}{1-r}\right) \end{array}\right), \label{eq:next_gen_matrix} \end{equation} where $\langle k_{s}\rangle$, $\langle k_{t}\rangle$ and $\langle k_{d}\rangle$ denote the expected static line stub, static triangle \emph{corner} and dynamic line stub degrees, $\langle s\rangle$, $\langle t\rangle$ and $\langle d\rangle$ denote the expected \emph{excess} static line stub, static triangle \emph{corner} and dynamic line stub degrees, and $r=\frac{\eta\beta_{d}}{(\eta+\gamma)(\beta_{d}+\gamma)}$. The basic reproduction number $R_{0}$ is the eigenvalue of the next generation matrix \eqref{eq:next_gen_matrix} with greatest magnitude. 

\subsection*{Model implementation}

A variable-order stiff differential equation solver (\emph{ode15s} in the MATLAB environment) was used to solve all relevant systems of equations. Initial conditions were specified, consisting of appropriate degree distributions and parameters for each edge-based compartmental model type, and of a user specified end time for the computation. 

Solutions to equations \eqref{eq:pgf_g}-\eqref{eq:R} were found using both interdependent and independent distributions for the three edge types. For interdependent distributions, a single probability distribution governed the distribution of \emph{pairs} of edge stubs, and additional model parameters $(p_{s}+p_{t}+p_{d})\equiv1$ were used to distribute each pair of stubs into: two static line stubs (with probability $p_{s}$), a single static triangle corner (with probability $p_{t}$), or two dynamic line stubs (with probability $p_{d}$). In such cases we used a negative binomial distribution for pairs of edge stubs with parameters $p$ and $r$ describing the probability of success in a single trial and the number of trial successes respectively, where the distribution itself is generated by $g_{nb}(x;r,p)=(\frac{p}{1-(1-p)x})^r$ and models the number of failures before a specified number of successes is reached in a series of identical, independent Bernoulli trials. We also utilised a discrete homogeneous distribution for pairs of edge stubs where all individuals had identical degree. For independent distributions, we used three separate Binomial distributions for the number of static line stubs, static triangle corners and dynamic line stubs.

\subsection*{Simulation implementation}

To test the validity of solutions to equations \eqref{eq:pgf_g}-\eqref{eq:R}, found in the MATLAB environment, Gillespie simulations \cite{gillespie1976general} were implemented to produce statistically-correct trajectories of SIR epidemic processes occurring on equivalent static-dynamic multiplex networks. Prior to each simulation, static and `dynamic' adjacency matrices were generated according to a configuration model approach, described as follows: for a population of $N$ individuals, three vectors of length $N$ are generated to record the number of static line stubs, static triangle corners, and dynamic line stubs associated with each individual, according to user-specified degree distributions provided to the script. The script ensures that the total number of static line stubs is even, the total number of dynamic line stubs is even, and that the total number of static triangle corners is a multiple of three. 

Firstly, the static network layer is generated using vectors containing the number of static line stubs and triangle corners each individual partakes in. Pairs of static line stubs and triples of static triangle corners are selected at random. Provided potential static lines and triangles do not generate self-loops (where an individual is joined to itself with an edge) or double edges (where an edge exists more than once within the static network layer), they are added to the static adjacency matrix. The \emph{unmatched} static line stubs and static triangle corners lists are updated, and the process continues until all static line stubs and triangle corners are successfully matched.

Secondly, the initial structure of the dynamic network layer is generated using the vector storing the number of dynamic line stubs each individual partakes in. Pairs of dynamic line stubs are selected at random. Provided a potential dynamic edge does not generate a self-loop or a double-edge within the dynamic network layer, it is added to the dynamic adjacency matrix. Successfully paired dynamic stubs are removed from the \emph{unmatched} stubs list, and the process continues until all dynamic line stubs are successfully matched. 

The nature of this configuration model approach means the wiring processes for the static and dynamic network layers may have to be restarted multiple times in order to achieve final network structures. Once all static line stubs, static triangle corners and dynamic line stubs have been wired up, the configuration process is complete. Although the script prevents double-edges from occurring within each network layer, it is possible for double-edges to occur \emph{across} the network layers, i.e. for two individuals to share both a static and a dynamic connection simultaneously. 

Given static and dynamic adjacency matrices describing the multiplex network structure, simulated epidemic processes can be implemented. In each Gillespie simulation, $\rho N$ initially infectious individuals are selected at random from the population. At each time step, a vector of length $(N+1)$ describes the state transition rate (infection or recovery) for all $N$ individuals, followed by a single edge swapping rate, $\frac{\eta M}{2}$, where $M:=$ total number of edges in the dynamic network layer. Inter-event times follow an exponential distribution with scale parameter $\frac{1}{R}$, where $R:=$ the sum of the rates vector at the current time step. Each event occurring is either an infection, a recovery or an edge swap. Uniformly distributed random numbers are generated at each time step to determine the next event to occur. When an edge swap event occurs, the script selects two dynamic edges at random, ensuring that all four nodes involved in these edges are unique. The script also ensures that the proposed new dynamic edges do not already exist within the dynamic network layer. Given these conditions, an edge swap occurs and the Gillespie process continues. The process terminates once the user specified end time is reached.

\section*{Results}

We have followed the EBCM approach to derive equations \eqref{eq:pgf_g}-\eqref{eq:R}, which describe an SIR epidemic spreading on a multiplex network consisting of two layers: a static network layer representing persistent human connections and a dynamic network layer representing temporary human interactions made outside of a typical household. The number of model equations remains fixed regardless of population size. We designed the multiplex model to afford control of network transitivity (clustering), on the static layer only, by generating the associated network structure using a combination of 2-vertex and 3-vertex cliques, referred to here as static lines and triangles. The dynamic network layer was generated via a single distribution for 2-vertex cliques. We have also applied the next generation matrix method \cite{diekmann2009construction} to compute the basic reproduction number $R_{0}$, a measure of the expected number of infections a typical infectious individual will cause during an epidemic. 

In what follows, we assess the validity of equations \eqref{eq:pgf_g}-\eqref{eq:R} and of the basic reproduction number $R_{0}$, obtained via the next generation matrix \eqref{eq:next_gen_matrix}. We firstly consider two extreme cases of the multiplex model: when either the static or the dynamic network layers are negligible (close to zero). In such cases, we show that predictions made by equations \eqref{eq:pgf_g}-\eqref{eq:R} resolve to predictions made by existing uniplex EBCM equations. When the full multiplex model is considered, with static and dynamic network elements present, there exists no basis for comparison other than generating exact simulations of the epidemic process. To this end, we utilise Gillespie simulations to demonstrate the validity of equations \eqref{eq:pgf_g}-\eqref{eq:R} in predicting the epidemic process for a number of multiplex network configurations. By solely considering the predictions of equations \eqref{eq:pgf_g}-\eqref{eq:R}, we explore the consequences of varying individual model parameters and of considering various combinations of model parameters $(p_{s}+p_{t}+p_{d})\equiv1$, governing the contributions of each edge type. Further, we explore the contributions of each edge type and how the resulting final epidemic size is altered within a systematic consideration of combinations of model parameters $\beta_{s}$, $\beta_{d}$ and $\eta$. Finally, we test the performance of the derived basic reproduction number $R_{0}$ in predicting the outcome of an epidemic and we explore variations in the value of $R_{0}$ and the associated final epidemic size predicted by equations \eqref{eq:pgf_g}-\eqref{eq:R} when altering the rate of rewiring, the extent of clustering, and the average degree in the multiplex model.

\subsection*{Model convergence to existing uniplex model equations}

\paragraph{Model without dynamic layer}

When the dynamic component of the dual-layer static-dynamic multiplex is removed, the model reduces to describe an SIR epidemic on a static uniplex network generated by lines and triangles. Biologically speaking, this reduced model tracks the epidemic as it spreads across persistent connections in a population with community structure. The EBCM approach has been followed to derive equations describing an SIR epidemic on such a network \cite{volz2011effects}. 

\begin{figure}
\centering
\includegraphics[width=\textwidth]{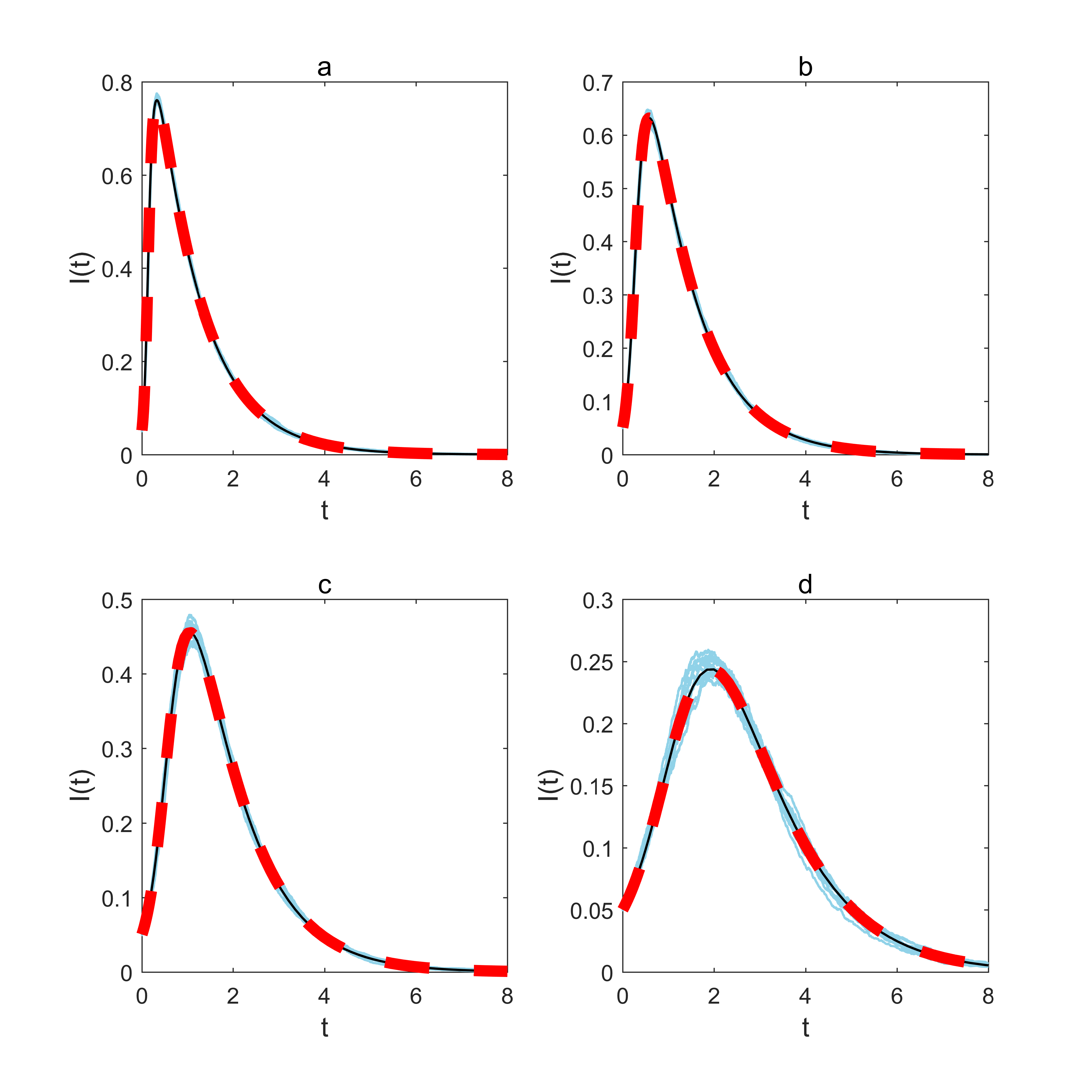}
\caption{{\bf Multiplex model convergence --- no dynamic layer, with simulation.} The time evolution of infection prevalence for the original EBCM of an SIR epidemic on a static uniplex network (solid black line), for the proposed EBCM of an SIR epidemic on a dual-layer multiplex with the dynamic network layer being close to zero (thick dashed red line), and for 10 Gillespie simulations of the SIR epidemic on a single network of size $N=5000$ (solid blue lines). In all panels $\gamma=1$, $\rho=0.05$, and $p=0.5$ and $r=10$ generate a negative binomial distribution for pairs of edge stubs. For the original static derivation (solid black line) $p_{s}=0.5=p_{t}$, describing the proportion of edge-pairs that are split into two single lines or remain as a triangle corner, respectively. For the multiplex derivation (thick dashed red line) $\beta_{s}=\beta_{d}$, $\eta=0.01$, and $p_{s}=0.4999999$, $p_{t}=0.5$ and hence $p_{d}=10^{-7}$ describe the proportion of edge-pairs that become two static lines, a static triangle corner, or two dynamic edges respectively. (a) $\beta 's=1$, $C=0.02677$, (b) $\beta 's=0.5$, $C=0.02670$, (c) $\beta 's=0.25$, $C=0.02658$, (d) $\beta 's=0.125$, $C=0.02685$, where $C$ denotes the global clustering coefficient of each static network layer generated for simulation}  \label{fig:modelconvergence_nodynamiclayer_varyingbeta_withsim}
\end{figure}

By comparing predictions made by uniplex model equations in \cite{volz2011effects} with those of multiplex model equations \eqref{eq:pgf_g}-\eqref{eq:R} when dynamic network elements are close to zero, we were able to test the multiplex model's convergence (Fig \ref{fig:modelconvergence_nodynamiclayer_varyingbeta_withsim}). Excellent agreement was observed between multiplex model equations where dynamic network elements are negligible, uniplex model equations \cite{volz2011effects} and Gillespie simulated epidemics on equivalent multiplex networks, for a number of scenarios with varying forces of infection.

\paragraph{Model without static layer}

When the static component of the dual-layer static-dynamic multiplex is removed, the model describes an SIR epidemic on a dynamic uniplex network generated by lines, where edges rewire at constant rate $\eta$ and degrees are conserved. Biologically, the reduced model describes an epidemic spreading through a population where connections between pairs of individuals are temporary but the number of connections an individual partakes in remains fixed. The EBCM approach was followed to derive equations describing an SIR epidemic spreading on such a network in \cite{miller2012edge}. 

\begin{figure}
\centering
\includegraphics[width=\textwidth]{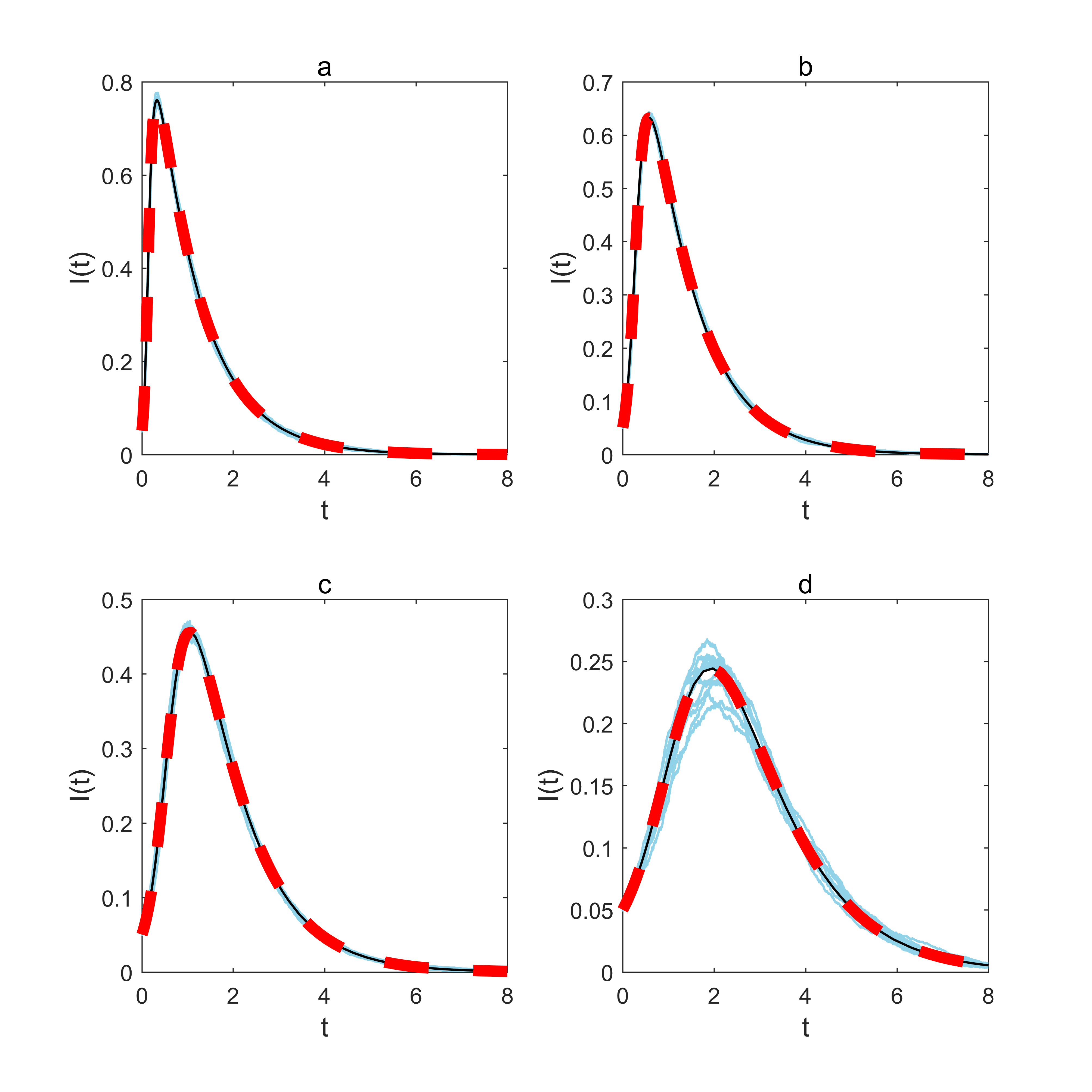}
\caption{{\bf Multiplex model convergence --- no static layer, with simulation.} The time evolution of infection prevalence for the original EBCM of an SIR epidemic on a dynamic uniplex network with conserved degrees and edge re-wiring (solid black line), for the proposed multiplex EBCM of an SIR epidemic with the static network layer being close to zero (thick dashed red line), and for 10 Gillespie simulations of the process on a single network of size $N=5000$ (solid blue lines). In all panels $\gamma=1$, $\rho=0.05$, and $p=0.5$ and $r=10$ generate a negative binomial distribution for pairs of edge stubs. For the original conserved-degree derivation (solid black line) $p_{d}=1$, indicating that all edge-pairs become two disjoint dynamic edges. For the multiplex derivation (thick dashed red line), $\eta=0.01$ and $p_{s}=p_{t}=10^{-7}$ and $p_{d}=0.9999998$ describe the proportion of edge-pairs that become two static lines, single triangle corners, or two dynamic edges respectively. (a) $\beta 's=1$, $C=0.004944$, (b) $\beta 's=0.5$, $C=0.005285$, (c) $\beta 's=0.25$, $C=0.005344$, (d) $\beta 's=0.125$, $C=0.005127$, where $C$ denotes the global clustering coefficient of each dynamic network layer generated for simulation, at time zero} \label{fig:modelconvergence_nostaticlayer_varyingbeta_withsim}
\end{figure}

Excellent agreement was observed between predictions made by equations \eqref{eq:pgf_g}-\eqref{eq:R} when static network elements are close to zero, output from the dynamic fixed-degree derivation in \cite{miller2012edge} and Gillespie simulations describing the SIR epidemic and edge rewire processes occurring simultaneously on equivalent multiplex networks, for a number of setups with varying forces of infection (Fig \ref{fig:modelconvergence_nostaticlayer_varyingbeta_withsim}).

\subsection*{Model validation by comparison with simulation}

We have observed excellent agreement between multiplex model predictions, uniplex model predictions and Gillespie simulated epidemics in extreme cases where either static or dynamic network elements are negligible (Figs \ref{fig:modelconvergence_nodynamiclayer_varyingbeta_withsim}-\ref{fig:modelconvergence_nostaticlayer_varyingbeta_withsim}). When multiplex network elements are non-negligible, static and dynamic network layers coexist in the model. In such cases, Gillespie simulated epidemics become the sole basis for assessing the validity of multiplex model equations \eqref{eq:pgf_g}-\eqref{eq:R}.

\begin{figure}
\centering
\includegraphics[width=\textwidth]{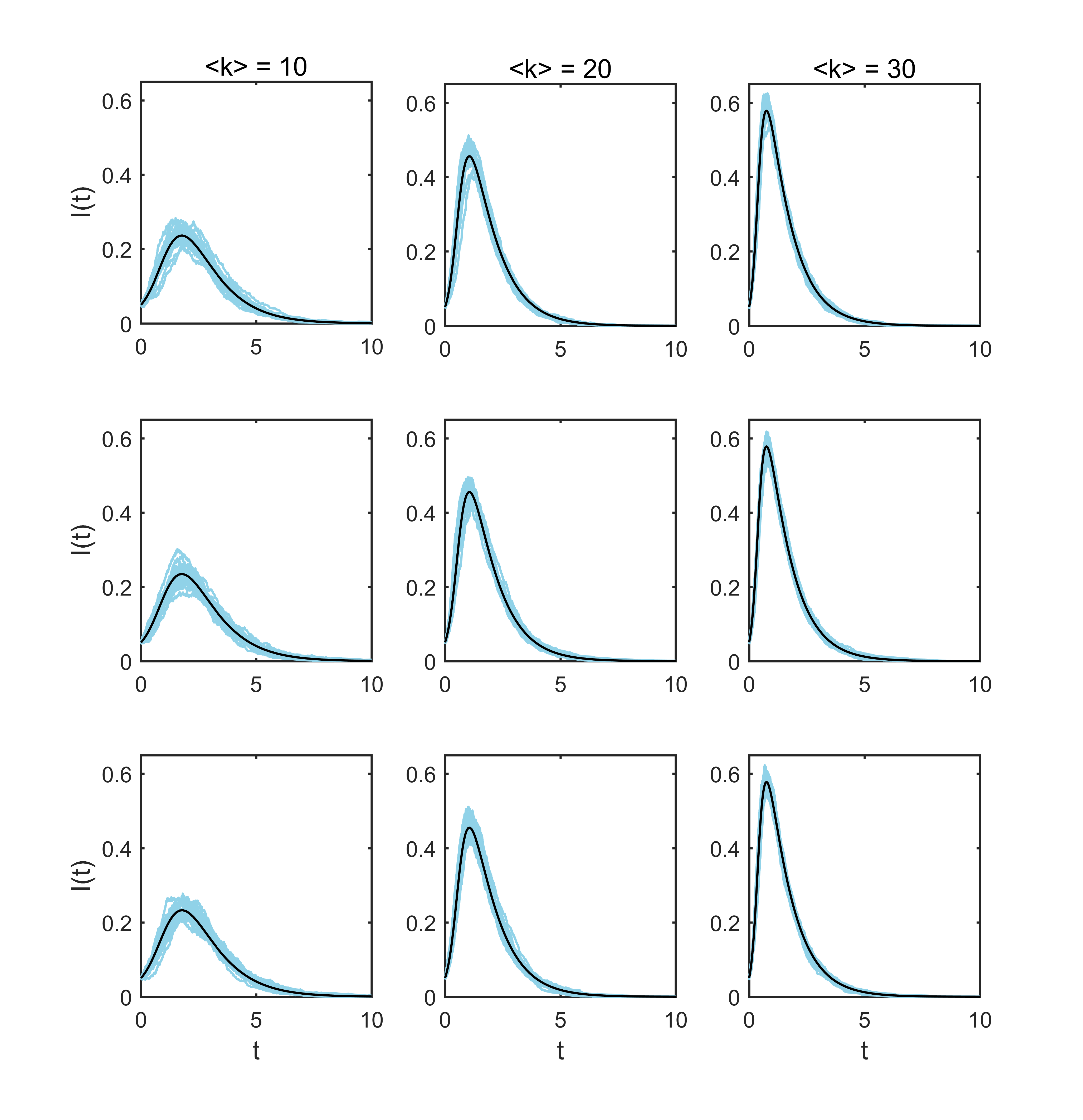}
\caption{{\bf Multiplex model prediction vs. simulation --- varying clustering and average degree.} Plotting the dynamics of the proportion of infected individuals over time. Each panel contains 25 Gillespie simulations on a single multiplex network comprised of $N=1000$ individuals (blue lines) and the associated EBCM prediction (black line). All networks are generated using a negative binomial distribution for pairs of edge stubs with parameters $p=0.5$ and various values for $r$. Networks in column 1 (counting from left to right) have average degree 10 (achieved via $r=5$), networks in column 2 have average degree 20 (achieved via $r=10$), and networks in column 3 have average degree 30 (achieved via $r=15$). Networks in row 1 (counting from top to bottom) have minimised clustering via values $p_{s}=0.99999998$ and $p_{t}=10^{-8}$. Networks in row 2 have the values $p_{s}=0.49999999=p_{t}$. Networks in row 3 have maximised clustering via the values $p_{s}=10^{-8}$ and $p_{t}=0.99999998$. Counting panels from left to right and top to bottom, starting with the upper-left panel, static networks have the following clustering coefficients: $C=0.0161$, $C=0.0267$, $C=0.0370$, $C=0.0535$, $C=0.0473$, $C=0.0493$, $C=0.0898$, $C=0.0662$, $C=0.0629$. In all panels, $tmax=10$, $\rho=0.05$, $\beta_{s}=\beta_{d}=0.25$, $\gamma=1$, $\eta=0.01$} \label{fig:panel_clust_vs_degree_sim_vs_full_model}
\end{figure}

\begin{figure}
\centering
\includegraphics[width=\textwidth]{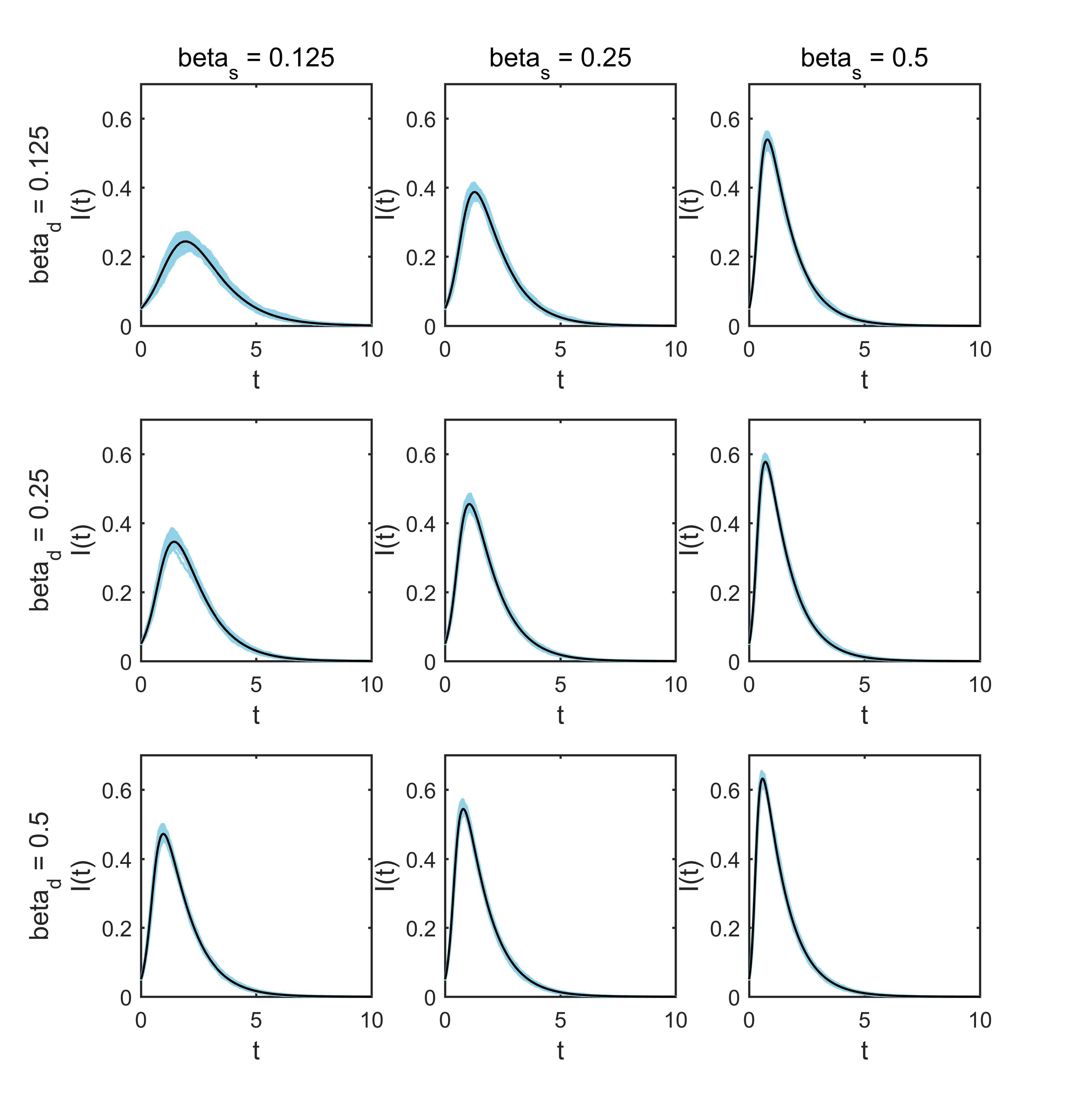}
\caption{{\bf Multiplex model prediction vs. simulation --- varying infection parameters $\boldsymbol{\beta_{s}}$ and $\boldsymbol{\beta_{d}}$.} Plotting the dynamics of the proportion of infected individuals over time. Each panel contains 100 Gillespie simulations (10 simulations on 10 multiplex networks comprised of $N=5000$ individuals) (blue lines) and the associated EBCM prediction (black line). All multiplex networks follow a negative binomial distribution for pairs of edge stubs with parameters $p=0.5$ and $r=10$, which were split into three edge types via $p_{s}=0.3=p_{t}$ and thus $p_{d}=0.4$. In all panels $tmax=10$, $\rho=0.05$, $\gamma=1$, $\eta=0.01$. Across the panels, different values for $\beta_{s}$ and $\beta_{d}$ have been used in the range $[0.125,0.25,0.5]$, indicated by individual column and row headings} \label{fig:panel_varying_betas_sim_vs_full_model_average_deg_10}
\end{figure}

A number of comparisons have been made between multiplex model predictions and Gillespie simulations when static and dynamic network elements coexist (Figs \ref{fig:panel_clust_vs_degree_sim_vs_full_model}-\ref{fig:panel_varying_betas_sim_vs_full_model_average_deg_10}). Excellent agreement was observed for a number of comparisons with various average degrees (imposed via negative binomial parameters $p$ and $r$, describing the distribution governing pairs of edge stubs) and various levels of clustering (imposed by varying parameter $p_{t}$ with the constraint $(p_{s}+p_{t}+p_{d})\equiv 1$) (Fig \ref{fig:panel_clust_vs_degree_sim_vs_full_model}). Excellent agreement was also observed for a number of comparisons with various combinations of the multiplex model's infection parameters $\beta_{s}$ and $\beta_{d}$ (Fig \ref{fig:panel_varying_betas_sim_vs_full_model_average_deg_10}).

\subsection*{A brief exploration of parameter spaces}

\begin{figure}
\centering
\includegraphics[width=\textwidth]{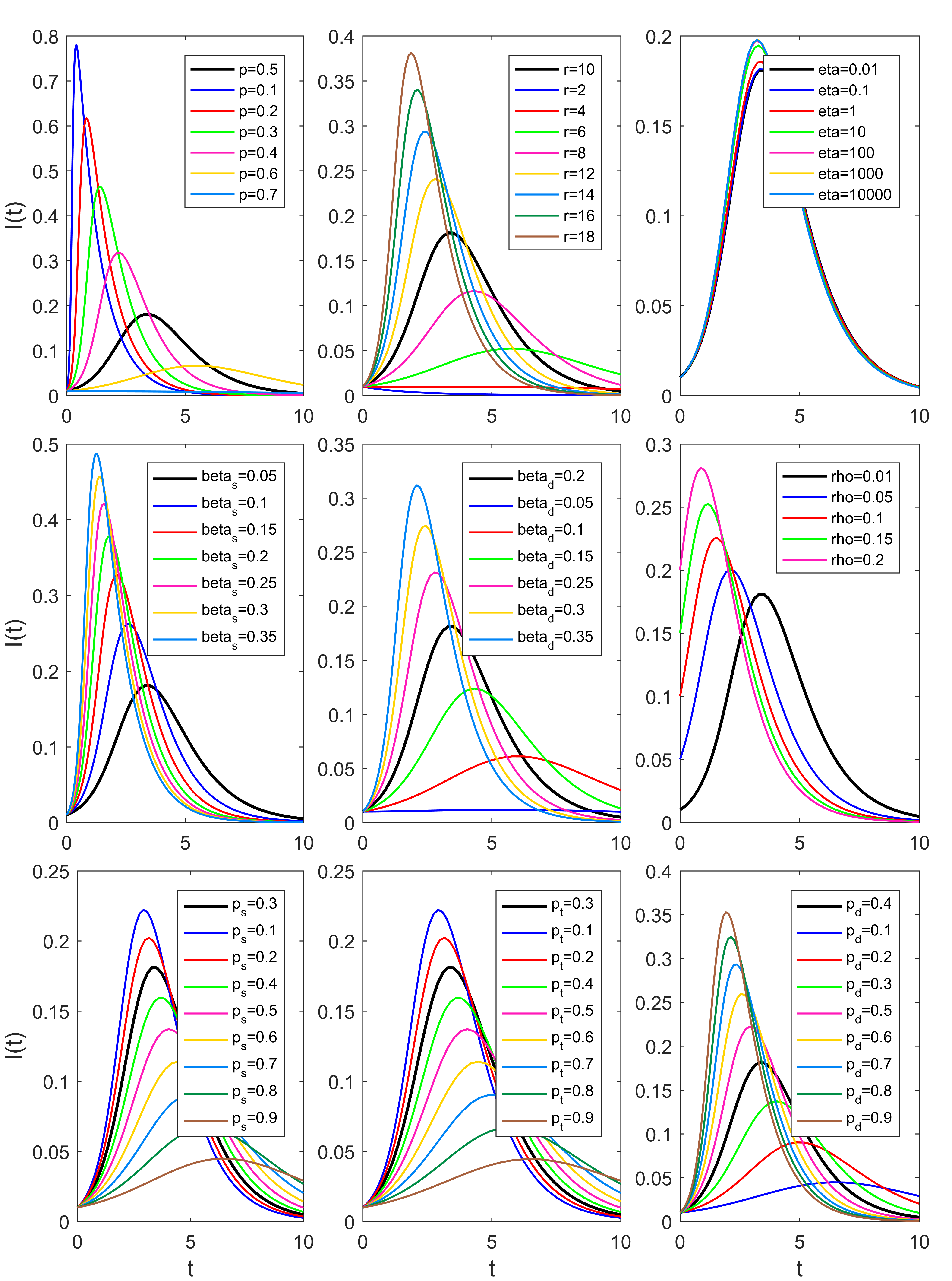}
\caption{{\bf Multiplex model predictions.} Plotting the dynamics of the proportion of infected individuals over time, for a number of different parameter sets. In all panels, a baseline parameter set ($p=0.5$, $r=10$, $p_s=0.3=p_t$, $p_d=0.4$, $\beta_{s}=0.05$, $\beta_{d}=0.2$, $\gamma=1$, $\eta=0.01=\rho$, $tmax=10$ $\Rightarrow$ $R_{0}=1.076$) is used to plot dynamics predicted by multiplex model equations \eqref{eq:pgf_g}-\eqref{eq:R} (thick black line). In each panel, a single parameter is varied and the resultant predictions are plotted in various colours, indicated by individual panel legends. In the bottom row of panels, parameters $p_{s}$, $p_{t}$ and $p_{d}$ are being varied. Since the model has the constraint $(p_{s}+p_{t}+p_{d})\equiv 1$, we alter the triplet values in each panel in the following way. Assume we are varying the parameter $p_{s}$. If the new $p_{s}$ is larger than the baseline $p_{s}$, we subtract $\tfrac{1}{2}$ the difference from the remaining baseline parameters $p_{t}$ and $p_{d}$. Conversely, if the new $p_{s}$ is smaller than the baseline $p_{s}$, $\tfrac{1}{2}$ the difference is added to each of the values $p_{t}$ and $p_{d}$} \label{fig:panel_varying_all_parameters}
\end{figure}

Having observed excellent agreement between simulated epidemic processes and equivalent predictions made by multiplex model equations, we investigated the effects of varying single parameters on the dynamics of epidemics predicted by equations \eqref{eq:pgf_g}-\eqref{eq:R}. In total, 9 individual model parameters were varied systematically whilst all (or the majority of) other parameters were held constant (Fig \ref{fig:panel_varying_all_parameters}). Across all parameters being varied, an identical baseline parameter set was utilised, with the resulting prediction made by equations \eqref{eq:pgf_g}-\eqref{eq:R} plotted in black to enable ease of comparison between different parameter scenarios.

This brief exploration highlights the effect that increasing or decreasing a single parameter has on the global dynamics of an SIR epidemic spreading across a dual-layer static-dynamic multiplex. Larger values of $p$ generate a negative binomial distribution with smaller average degree and a reduction in variance, slowing the epidemic's spread. Larger values of $r$ led the epidemic to spread more rapidly due to an increase in average degree and variance of the negative binomial distribution for pairs of edge stubs. Varying the rewiring rate $\eta$ led to less pronounced differences, where larger values of $\eta$ led to a slight increase in the speed at which the epidemic spread through the population. Increasing a single infection parameter $\beta_{s}$ or $\beta_{d}$ leads to an increase in the rate of epidemic spread. Altering the parameter $\rho$ means changing the number of individuals who are infectious at the start of an epidemic process. Increasing the value of $\rho$ leads to changes in the shape of the curve $I(t)$, describing the prevalence of infection at time $t$, and to the epidemic process finishing sooner. Altering the values $p_{s}$, $p_{t}$ and $p_{d}$, with the constraint $(p_{s}+p_{t}+p_{d})\equiv1$, demonstrates the range of dynamics that can be achieved using a fixed distribution for pairs of edge stubs with additional parameters to distribute edge pairs into three edge types. Baseline infection parameters are used across all three panels, thus $\beta_{s}=0.05<0.2=\beta_{d}$, meaning that an increase in the proportion of dynamic edges leads to an increase in the speed of the epidemic, whilst any increase in the proportion of static edges leads to a decrease in the rate of epidemic spread.

\subsection*{Contribution of network layers via $\boldsymbol{(p_{s}+p_{t}+p_{d})\equiv1}$}

When degree distributions are interdependent, the parameters $(p_{s}+p_{t}+p_{d})\equiv1$ afford the ability to investigate the effects on epidemic dynamics of altering the proportion of edges of each type. Previously, we observed changes in the dynamics of $I(t)$, caused by altering the contributions of each edge type (Fig \ref{fig:panel_varying_all_parameters}), where $\beta_{d}>\beta_{s}$, rewiring was slow, and pairs of edge stubs were governed by a negative binomial distribution.

\begin{figure}
\centering
\includegraphics[width=\textwidth]{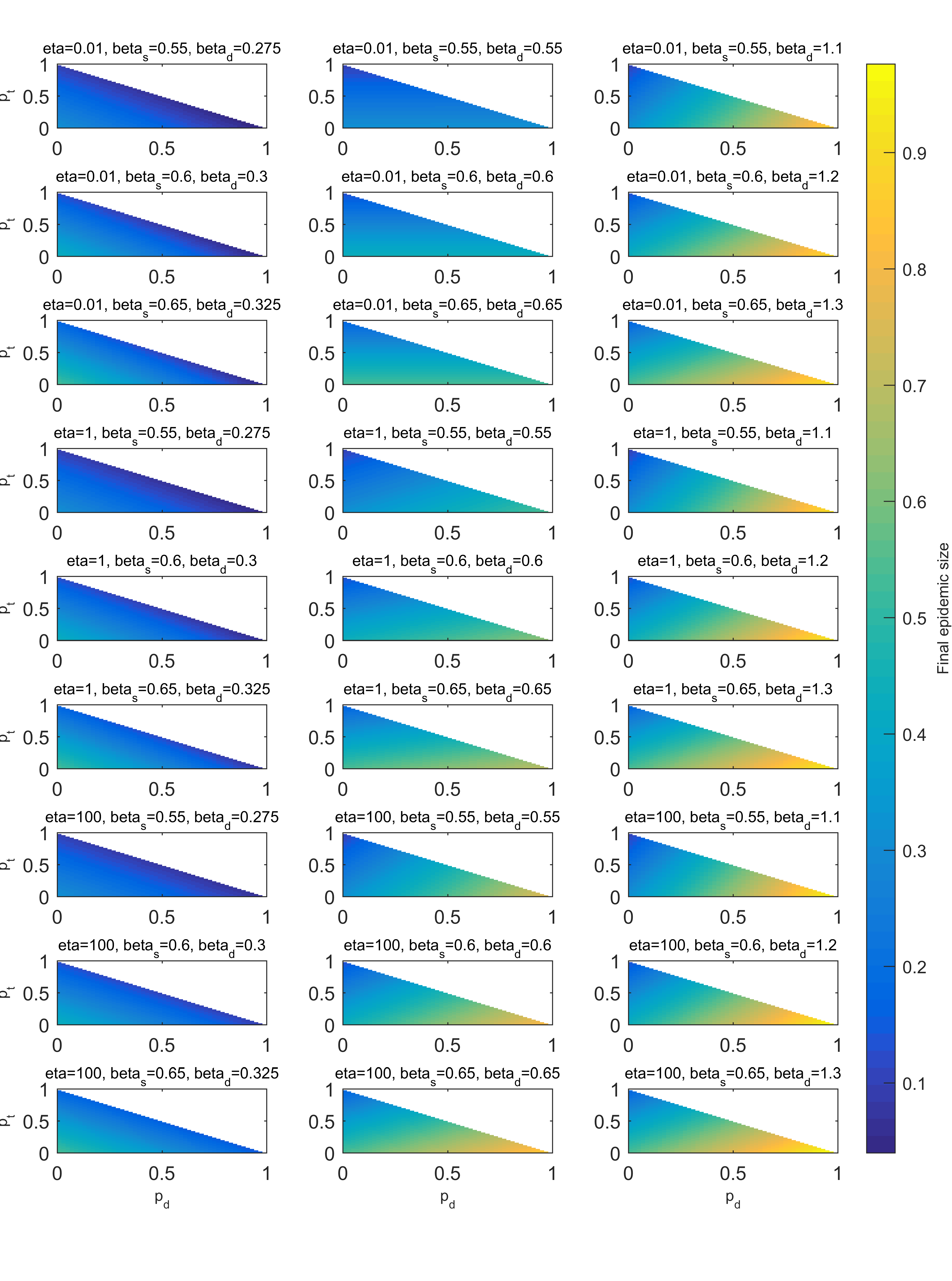}
\caption{{\bf Multiplex model layer contributions.} Heat map plots depicting the final epidemic size (equal to the fraction of the population who are either infectious or recovered at the end of the epidemic process) predicted by equations \eqref{eq:pgf_g}-\eqref{eq:R} for a multiplex network of various proportions $p_{s}$, $p_{t}$ ($y$-axis) and $p_{d}$ ($x$-axis), with the model constraint $(p_{s}+p_{t}+p_{d})\equiv1$. For all setups $\gamma=1$, $\rho=0.01$, $tmax=25$ and pairs of edge stubs followed a discrete homogeneous distribution where all individuals had 2 edge pairs (and hence total degree 4). The values of remaining model parameters $\eta$, $\beta_{s}$ and $\beta_{d}$ are indicated above each panel, with $\eta \in [0.01,1,100]$, $\beta_{s}\in [0.55,0.6,0.65]$ and $\beta_{d}\in [\beta_{s}/2, \beta_{s},2\beta_{s}]$. All 27 possible combinations of the parameters $\eta$, $\beta_{s}$ and $\beta_{d}$ are considered. Prior to implementation, a number of set-ups across the $(p_{s},p_{t},p_{d})$ parameter spaces in each panel were tested by hand to ensure that the epidemic process had concluded by time $tmax=25$} \label{fig:heatplot_layer_contributions}
\end{figure}

In this multiplex setting, increasing the force of infection on one network layer effectively reduces the force of infection on remaining network layers. Thus the value of parameters $\beta_{s}$, $\beta_{d}$ and $\eta$, and the ratios between them, bias the effect of varying model parameters $p_{s}$, $p_{t}$ and $p_{d}$. To take this into account, we allowed parameters $\beta_{s}$, $\beta_{d}$ and $\eta$ to take three distinct values (specifically $\beta_{s}\in [0.55,0.6,0.65]$, $\beta_{d}\in [\frac{\beta_{s}}{2},\beta_{s},2\beta_{s}]$ and $\eta\in[0.01,1,100]$), and we considered all 27 combinations of their values, before varying the contributions of each edge type and recording the final epidemic size predicted by equations \eqref{eq:pgf_g}-\eqref{eq:R} in each case (Fig \ref{fig:heatplot_layer_contributions}). This approach enabled isolation of the effects of changing single infection or rewiring parameters and exploration of the contributions made by various combinations of edge proportions $p_{s}$, $p_{t}$ and $p_{d}$ in distinct parameter settings.

Increasing the proportion of triangle corners via $p_{t}$ consistently led to decreases in final epidemic size, suggesting that clustering slows the epidemic process regardless of the choice of parameters $\beta_{s}$, $\beta_{d}$ and $\eta$ (Fig \ref{fig:heatplot_layer_contributions}). Generally, increasing the value of $\eta$ resulted in an increase in final epidemic size when comparing identical edge contributions. Likewise, increasing the value of infection parameters $\beta_{s}$ or $\beta_{d}$ led to an increase in final epidemic size. Dependant on the combination of parameters $\beta_{s}$, $\beta_{d}$ and $\eta$, different behavioural regimes emerge, indicated by the orientation of colours and the direction in which they change in individual panels. We observe that a single edge proportion can have a more or less dominant effect on the outcome, dependent on the particular parameter set. For example, when $\eta=0.01$ and $\beta_{s}=0.55=\beta_{d}$, changing the proportion of dynamic edges $p_{d}$ has little effect on the final epidemic size. However, when $\eta=100$, $\beta_{s}=0.65$ and $\beta_{d}=1.3$, altering the parameter $p_{d}$ leads to more extreme changes in final epidemic size, a result of $\beta_{d}$ dominating $\beta_{s}$ and an increased rate of dynamic edge rewiring.

\subsection*{Validation of basic reproduction number $\boldsymbol{R_{0}}$}

The next generation matrix $\boldsymbol{G}$ \eqref{eq:next_gen_matrix} and the value $R_{0}$ can be validated by testing to see if the final epidemic size is disturbed as $R_{0}$ exceeds the epidemic threshold ($R_{0}=1$). When the basic reproduction number is sub-threshold ($R_{0}<1$), the associated epidemic process is expected to `die-out'. However, when $R_{0}>1$ the epidemic is expected to take hold and spread within a population. 

\begin{figure}
\centering
\includegraphics[width=0.8\textwidth]{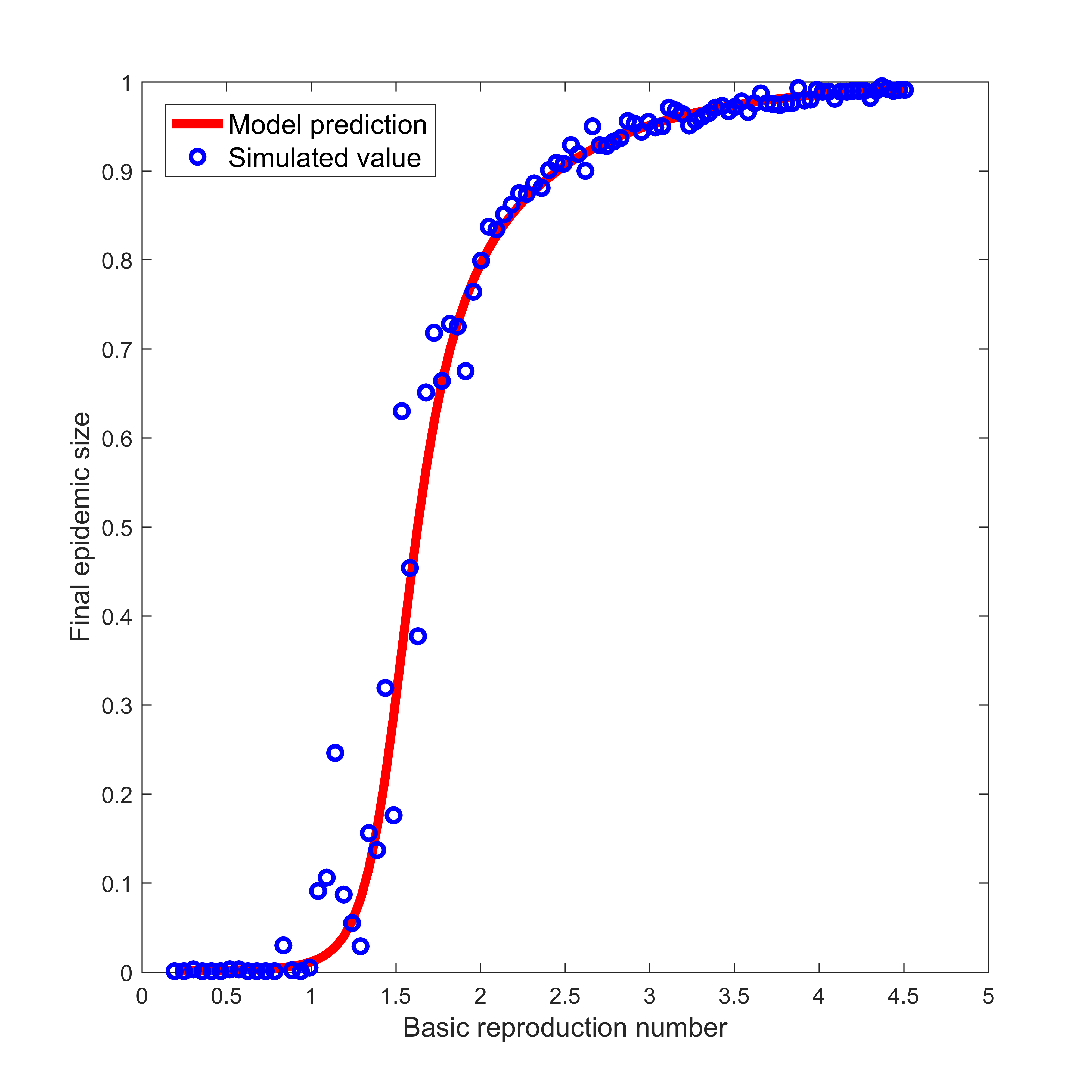}
\caption{{\bf Validation of the basic reproduction number $\boldsymbol{R_{0}}$.} Plotting values of the basic reproduction number $R_0$ ($x$-axis), found via the leading eigenvalue of the matrix \eqref{eq:next_gen_matrix}, against the associated final epidemic sizes ($y$-axis) predicted by multiplex equations \eqref{eq:pgf_g}-\eqref{eq:R} (red line) and recorded by single statistically-correct Gillespie simulations (blue circles). Static and dynamic line stubs follow Binomial distributions with parameters $n=20$ and $p=0.5$. The distribution of triangle corners follows a Binomial distribution with parameters $n=1$ and $p=0.001$ to minimise clustering. Fixed parameters were $\gamma=1$, $\rho=0.001$, $\eta=0.01$, $tmax=10$, $N=1000$. In each setup $\beta_{s}=\beta_{d}$. 100 transmission rates were tested, from $\beta_{s}=\beta_{d}=0.01$ up to $\beta_{s}=\beta_{d}=0.3$, in equal-sized increments. In Gillespie simulations where $R_0 > 1$, if the number of infected individuals did not reach 10 times the initial number of infectives, all data was discarded and the Gillespie script restarted from initial conditions at time zero} \label{fig:validation_of_R0_6}
\end{figure}

For a number of set-ups, we recorded the final epidemic size predicted by equations \eqref{eq:pgf_g}-\eqref{eq:R}, the final epidemic size of a single Gillespie simulation of the same process, and the associated $R_{0}$ value (Fig \ref{fig:validation_of_R0_6}). To obtain a suitable range of $R_{0}$ values we systematically increased $\beta_{s}=\beta_{d}$ from sub-threshold values, while all other parameters were held constant. Independent Binomial distributions were used for static line stubs, static triangle corners, and dynamic line stubs. In Gillespie simulations where $R_{0}>1$, we imposed an additional constraint requiring the number of infectives to reach at least ten times the initial number of infected individuals, otherwise a new Gillespie simulation was implemented. As $R_{0}$ exceeded the epidemic threshold, the final epidemic size predicted by model equations \eqref{eq:pgf_g}-\eqref{eq:R} and from individual simulations increased rapidly, suggesting the derivation of the next generation matrix $\boldsymbol{G}$ and associated $R_{0}$ is rigorous.

\begin{figure}
\centering
\includegraphics[width=\textwidth]{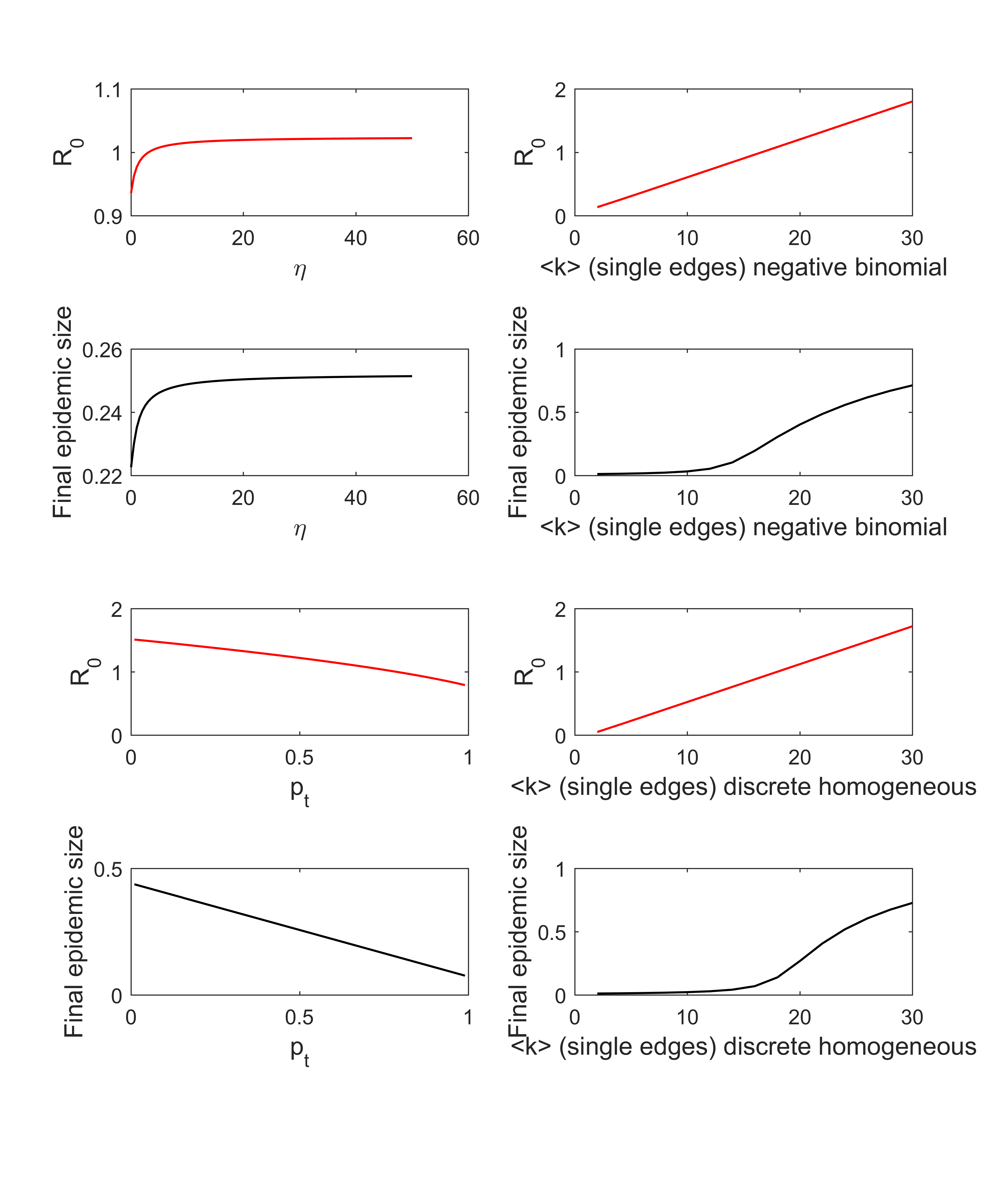}
\caption{{\bf{Effects of rewiring, average degrees and clustering.}} Plotting the value of $R_0$ and the associated final epidemic size found using EBCM equations \eqref{eq:pgf_g}-\eqref{eq:R}, for a number of different set-ups. Upper-left panels: Testing 100 evenly-spaced values for $\eta$ in the range $[0.01,50]$. Remaining model parameters were $p_{s}=0.3=p_{t}$, $\beta_{s}=0.1=\beta_{d}$, $\gamma=1$, $\rho=0.01$ and $tmax=25$. Pairs of edge stubs followed a negative binomial distribution with parameters $p=0.5$ and $r=5$. Upper-right panels: Testing 15 evenly-spaced values for $\langle k\rangle \in [2,30]$, generated using a negative binomial distribution for pairs of edge stubs with fixed $p=0.5$ and $r \in [1,15]$. Remaining model parameters were $p_{s}=0.3=p_{t}$, $\beta_{s}=0.0625=\beta_{d}$, $\gamma=1$, $\eta=0.1$, $\rho=0.01$, $tmax=25$. Lower-left panels: Testing 100 evenly-spaced values for $p_{t}$ in the range $[0.01,0.99]$. The proportion $(1-p_{t})$ was split equally between parameters $p_{s}$ and $p_{d}$. Remaining model parameters were $\beta_{s}=0.5=\beta_{d}$, $\gamma=1$, $\rho=0.01$, $\eta=0.1$ and $tmax=25$. Pairs of edge stubs followed a discrete homogeneous distribution where all individuals had 2 edge pairs. Lower-right panels: Testing 15 evenly-spaced values for $\langle k\rangle \in [2,30]$, generated using a discrete homogeneous distribution for pairs of edge stubs where all individuals have identical degree. Remaining model parameters were $p_{s}=0.3=p_{t}$, $\beta_{s}=0.0625=\beta_{d}$, $\gamma=1$, $\rho=0.01$, $\eta=0.1$, $tmax=25$}
\label{fig:exploration_of_R0}
\end{figure}

We plotted $R_{0}$ and the associated final epidemic size predicted by equations \eqref{eq:pgf_g}-\eqref{eq:R} for a number of scenarios to investigate the impact on their values of varying specific multiplex network attributes (rewiring, clustering and average degree), and to explore the relationship between $R_{0}$ and final epidemic size (Fig \ref{fig:exploration_of_R0}). Varying the rewiring rate $\eta$ demonstrates that $R_{0}$ and the associated final epidemic size increase with the value of $\eta$. Varying $\eta$ can also move the system below or above the epidemic threshold $R_{0}=1$. However, there is a limit to this relationship; as $\eta$ increases above 20, the changes in $R_{0}$ and final epidemic size are negligible. We have seen previously that larger values of $p_{t}$ result in smaller final epidemic sizes, suggesting that increased clustering slows epidemic processes on multiplex networks (Fig \ref{fig:heatplot_layer_contributions}). Here, we find that increasing $p_{t}$ leads to decreases in both $R_{0}$ and the associated final epidemic size (Fig \ref{fig:exploration_of_R0}). The relationship between $p_{t}$ and final epidemic size appears to be linear. For smaller $p_{t}$ the curve with $R_{0}$ appears to be linear, but as $p_{t}$ tends towards its maximal value, the reduction in $R_{0}$ increases.

An increase in average degree $\langle k\rangle$, where pairs of edge stubs follow a negative binomial distribution, led to increases in $R_{0}$ and final epidemic size (Fig \ref{fig:exploration_of_R0}). The relationship between $\langle k\rangle$ (negative binomial) and $R_{0}$ appears to be linear. However, the relationship between $\langle k\rangle$ and final epidemic size differs. The final epidemic size increases at a faster rate above some critical average degree, say $\langle k\rangle=12$. A similar pattern emerges in the relationship between the average degree, $R_{0}$ and final epidemic size when pairs of edge stubs follow a discrete homogeneous distribution. Again, the final epidemic size begins to increase more quickly after some critical average degree has been reached, around $\langle k\rangle=14$. These results show that small average degrees make it hard for the epidemic to take hold in the population. Potentially, this is a result of the multiplex network becoming divided into more than one connected component, meaning the disease can get trapped within smaller sub-populations of individuals, limiting its effect.

\section*{Discussion}

We have proposed a model describing the time evolution of an SIR epidemic spreading through a population of individuals in a dual-layer static-dynamic multiplex network. The model incorporates heterogeneity in the structure, type and duration of connections between individuals. Following the EBCM approach \cite{miller2014epidemics}, we obtained expressions for time-evolving quantities of interest, such as the infectious proportion of the population $I(t)$. An estimate of the associated basic reproduction number $R_{0}$ was derived, utilising the next generation matrix method \cite{diekmann2009construction}. 

Multiplex model equations \eqref{eq:pgf_g}-\eqref{eq:R} were validated, first by testing convergence of epidemic dynamics to predictions made by existing uniplex edge-based compartmental model equations, when either network layer (static or dynamic) was eliminated, and second by comparing full model (with static and dynamic elements) predictions to the dynamics of corresponding statistically-correct Gillespie simulations \cite{gillespie1976general}. 

The multiplex model's parameter space was explored by varying individual parameters and plotting the resulting epidemic dynamics, and by mapping the outcome on final epidemic size of having various proportions of each edge type when considering different combinations of model parameters $\beta_{s}$, $\beta_{d}$ and $\eta$. The basic reproduction number $R_{0}$, found via the leading eigenvalue of the next generation matrix $\boldsymbol{G}$ \eqref{eq:next_gen_matrix}, was validated by demonstrating that continually incrementing infection parameters $\beta_{s}$ and $\beta_{d}$, with all else held constant, led to a rapid increase in final epidemic size as $R_{0}$ exceeded its epidemic threshold. Finally, we explored the effect on $R_{0}$ and the associated final epidemic size predicted by equations \eqref{eq:pgf_g}-\eqref{eq:R} of altering specific multiplex network attributes governing the rate of rewiring, the extent of clustering and the average degree.

Our unique contribution towards the literature is a model with a combination of static \emph{and} dynamic network elements, derived by combining the EBCM approach to modelling an SIR epidemic on a static network with tunable clustering \cite{volz2011effects} with the EBCM approach to modelling an SIR epidemic on a dynamic fixed-degree network \cite{miller2012edge}, under the framework of a dual-layer multiplex network. 

The EBCM approach allows us to model variations in contact structure, contact type, and contact duration simultaneously. Modelling such heterogeneities via EBCM provides an opportunity to investigate the effects of heterogeneities observed in real-world networks \cite{perry2003social,komurov2007revealing,vernon2009representing}, alongside consideration of common network attributes such as clustering and degree distributions. EBCM also affords a huge reduction in the number of equations required to track the epidemic, compared with full simulation.

This work progresses the drive to derive population models that capture reasonable levels of complexity and heterogeneity whilst exhibiting a tractable number of equations. By providing a clear and concise `walkthrough' to deriving and validating our desired model, we hope that future researchers are inspired to build on these results by designing and implementing novel models, modelling approaches, and computational algorithms.

The work here extends previous research following the edge-based compartmental modelling approach. Prior EBCM approaches derived model equations describing the SIR epidemic process on wholly static or wholly dynamic uniplex networks. For example, EBCM has been utilised to describe the SIR epidemic on static actual-degree configuration model (CM) networks \cite{miller2012edge}, static CM networks with tunable clustering \cite{volz2011effects}, and static expected degree mixed Poisson (MP) networks \cite{miller2012edge}. 

Dynamic uniplex networks have also been considered via the EBCM approach. Namely, CM networks with mean-field social heterogeneity (edges are broken and rewired at a very fast rate, meaning all pairs of individuals contact each other at the same rate, and edge durations are fleeting), dynamic fixed-degree CM networks (edges are rewired but edge durations are finite), dormant contact CM networks (existing edges are broken and remain dormant for some time, before being re-established), MP networks with mean-field social heterogeneity (fleeting edge duration), and dynamic variable-degree MP networks (finite edge duration) \cite{miller2012edge}.

Existing modelling approaches incorporating heterogeneity include the consideration of an epidemic with two `levels' of mixing between individuals (but no network structure) \cite{ball1997epidemics}, and the later considerations of epidemic processes occurring on structured populations with two levels of mixing \cite{zhang2015optimizing}, and with two routes of transmission \cite{zhao2014multiple}. Recently, the EBCM approach was used to derive equations describing an SIR epidemic process with non-sexual and sexual transmission routes, a characteristic of diseases such as Ebola and Zika \cite{miller2017mathematical}. 

Other modelling approaches have incorporated dynamicity of connections between individuals (and hence heterogeneity in contact duration) by, e.g., considering an SIR epidemic on a network with intermittent social distancing, where susceptible individuals break links with infectious individuals for some time $t_{b}$, after which the connection is re-established \cite{valdez2013temporal}. Another approach considered the effects of constrained rewiring during an SIS epidemic, whereby susceptible individuals cut links to infectious individuals regardless of distance, and rewire to a susceptible individual within a given radius, where the nodes of the network were embedded in Euclidean space \cite{rattana2014impact}. 

Research considering the large graph limit of an SIR epidemic on a dynamic multilayer network affords heterogeneity in contact type and in contact duration by allowing individual network layers to contain either activating or de-activating edges, and by allowing edges in different layers to correspond to different types of contacts \cite{jacobsen2016large}. Although Jacobsen et al. \cite{jacobsen2016large} consider the SIR epidemic spreading on a multiplex network, including providing a dual-layer multiplex example where edge types correspond to community and healthcare contacts, they do not consider any fully static network components.

There are a number of adaptations that can be made to the proposed model. The model considers a heterogeneous contact structure between $N$ individuals. However, the locations of $N$ individuals are not taken into account. Real-world networks occur in space \cite{barthelemy2011spatial} and thus it is important to investigate the effects of considering node locations. In this study, we have chosen to disregard the spatial locations of individuals. A more realistic model of an SIR epidemic spreading on a multiplex network of individuals would be achieved by embedding the locations of each individual into Euclidean space. Even more complex models could consider dynamic node locations, or a combination of static and dynamic node locations. 

Another potential adaptation is considering weighted network connections. In the proposed model, all connections are considered to be unweighted, or equivalently to share equal weight (homogeneity). The model could be adapted by, e.g., making the weight of each connection proportional to the Euclidean distance between the two node locations (given spatial embedding), by imposing a distribution of connection weights, or by assigning weights at random. Then, the probability of contracting disease across a connection can be made proportional to the weight of that connection. 

In the present model, the population of $N$ individuals is fixed. We do not consider the effect of flux in or out of the population, e.g. by births, deaths or migration events. An important next step is to adapt the model presented here to consider in- and outflow of members of the population, or at least to consider whether such in- and outflows significantly influence disease dynamics. Other model adaptations include allowing for tunable clustering on all network layers (and thus imposing two edge distributions on each network layer), implementing more complex distributions governing the degrees of each node, biasing initially infectious individuals instead of selecting them at random, and considering alternative rules for edge dynamicity, such as constrained rewiring \cite{rattana2014impact}, edge activation and deletion \cite{jacobsen2016large,selley2015dynamic,taylor2012epidemic}, or the dormant contact approach \cite{valdez2013temporal,shkarayev2014epidemics,tunc2013epidemics}.

The multiplex model affords tunable clustering on the static network layer by generating its contact structure using a distribution of line stubs and a distribution of triangle corners. However, the configuration model wiring process requires that any two individuals share at most one connection within a single network layer. Double edges can occur across network layers (i.e. when the same edge is present in both network layers), but not within them. This constraint greatly reduces the possibilities for placing triangles suitably into the network, meaning the configuration process is slowed down and the extent of clustering that can be achieved is reduced. Greater control over clustering could be achieved by adapting the model to allow for overlapping triangles (and either allowing double edges to occur in single network layers, or amalgamating any double edges that occur into single edges, or doubly-weighted edges).

Other than making adaptations to the proposed model, there are a number of tests and analyses which are beyond the scope of this work. Firstly, a comprehensive exploration of the entire parameter space would elucidate the behavioural `envelope' of the model and uncover any parameter regions where the model poorly predicts the SIR epidemic process, compared with simulation. A more thorough understanding of the impact of degree and degree heterogeneity on the relationship between parameters and system behaviour will require consideration of additional edge distributions with various levels of heterogeneity and average degrees. Secondly, the model's utility can be investigated by using real-world data from historical epidemics or similar processes, e.g. livestock herd contact tracing data or Twitter data tracking the prevalence of a hashtag over time. Using real data, model parameters could be estimated using Bayesian estimation techniques and the resulting model predictions compared with prior knowledge of what occurred. The basic reproduction number $R_{0}$ can be tested in the same way.

This work considers an SIR compartmental model under the guise of a disease spreading through a networked population. Thought must be given to what other real-world processes can be well described by the SIR compartmental model, such as opinion formation, rumour spreading or uptake of fashion trends. Further, a two-layer multiplex like the proposed model could be used to investigate the dynamics of two \emph{interacting} SIR-type processes, such as a physical disease spreading process occurring on one network layer in combination with a disease awareness process occurring on the opposing network layer, using similar approaches to those of \cite{funk2009spread} and \cite{li2015multiple}.

Future research can build on these observations by considering similar modelling approaches that account for compartmental models other than the SIR-type. For example, the SIS model (describing infections that do not confer lasting immunity, such as the common cold) and the SEIR model (describing infections with incubation periods, where individuals have contracted a disease but are not yet infectious and hence are in the `exposed' disease state) are not considered here. Modelling an SEIR infection may require simple adaptation of the existing EBCM approach. However, consideration of an SIS-type epidemic process requires an altogether new modelling approach. A key assumption of the present approach is the consideration of all neighbours of the test node $u$ as being independent. Attempting to impose this assumption would prevent modelling of SIS dynamics, a consequence which is discussed in \cite{miller2014epidemic} and \cite{miller2012edge}.

Experiments that can be performed to improve and inform future modelling approaches include: quantifying the levels of heterogeneity in existing populations, including behavioural and structural heterogeneity, gaining a deeper understanding of the biological processes underlying disease spreading processes, improving on existing algorithmic and analytic approaches, and fostering closer relations between modellers and practitioners, in order to maximise the benefits arising from research.

\section*{Acknowledgments}

RCB acknowledges funding from the Engineering and Physical Sciences Research Council, EP/M506667/1. JCM was funded by the Global Good Fund through the Institute for Disease Modeling. The funders had no role in study design, data collection and analysis, decision to publish, or preparation of the manuscript.

\bibliographystyle{plos2015}
\bibliography{bib_project2}

\end{document}